\documentclass[preprint]{aastex}

\usepackage{natbib}
\usepackage{lineno}

\usepackage{subcaption}
\usepackage{graphicx}
\usepackage{rotating}
\usepackage{epsfig}
\usepackage{epstopdf}
\usepackage{enumitem}
\usepackage{amssymb,amsmath}
\usepackage{xcolor}
\usepackage{chngpage}

\newcommand{\Rs}{$ R_{\odot}$}

\newcommand{\de}{$^{\circ}$}

\newcommand{\pB}{\textit{pB}}

\newcommand{\kms}{\,km$\,s^{-1}$}

\newcommand{\insitu}{\emph{in situ}}

\newcommand{\app}{$\approx$}
\makeatletter

\newcommand{\Rmnum}[1]{\expandafter\@slowromancap\romannumeral #1@}

\begin{document}

\shorttitle{Solar wind boundary condition based on density}
\shortauthors{Morgan}
\title{An inner boundary condition for solar wind models based on coronal density}
\author{Kaine A. Bunting}
\author{Huw Morgan}
\affil{Department of Physics, Aberystwyth University, Ceredigion, Cymru, SY23 3BZ, UK}
\email{kab84@aber.ac.uk}

\begin{abstract}
Accurate forecasting of the solar wind has grown in importance as society becomes increasingly dependent on technology that is susceptible to space weather events. This work describes an inner boundary condition for ambient solar wind models based on tomography maps of the coronal plasma density gained from coronagraph observations, providing a novel alternative to magnetic extrapolations. The tomographical density maps provide a direct constraint of the coronal structure at heliocentric distances of 4 to 8\Rs, thus avoiding the need to model the complex non-radial lower corona. An empirical inverse relationship converts densities to solar wind velocities which are used as an inner boundary condition by the Heliospheric Upwind Extrapolation (HUXt) model to give ambient solar wind velocity at Earth. The dynamic time warping (DTW) algorithm is used to quantify the agreement between tomography/HUXt output and \insitu\ data. An exhaustive search method is then used to adjust the lower boundary velocity range in order to optimize the model. Early results show up to a 32\% decrease in mean absolute error between the modelled and observed solar wind velocities compared to that of the coupled MAS/HUXt model. The use of density maps gained from tomography as an inner boundary constraint is thus a valid alternative to coronal magnetic models, and offers a significant advancement in the field given the availability of routine space-based coronagraph observations.

\end{abstract}

\maketitle

\vspace{0.5cm}
\section{Introduction}

{Rapid changes in solar wind conditions have a direct impact on Earth's magnetosphere \citep[e.g.,][]{MEZIANE20141}, ionosphere \citep[e.g.,][]{Milan2007}, and both ground-based and space-based technology \citep[e.g.,][]{Imken2018,Baker2004,Doherty2004}. Estimates of economic impact of space weather on European power grids alone range from \texteuro10s-100s billion \citep[e.g.,][]{eastwood2018}. Risk can be mitigated considerably by early warnings of impending space weather, currently undertaken by organisations such as the Meteorological Office in the UK. We believe that improvements in space weather forecasting depend primarily on three related categories: (1) Better observations of the Sun, corona, and solar wind; (2) Greater understanding of the physical processes operating in the corona and solar wind; (3) Improvements in data analysis and forecasting methods. This work presents a novel boundary condition to solar wind models based on tomography maps created from coronagraph observations of the solar atmosphere, thus an advancement that belongs to the third category, and can contribute to the second.

The use of tomography maps as an inner boundary condition to a solar wind model, combined with an ensemble approach, plays a key role in the Coronal Tomography (CorTom) module to the Space Weather Empirical Ensemble Package (SWEEP) project. SWEEP is funded by the UK government's Space Weather Instrumentation, Measurement, Modelling, and Risk (SWIMMR) scheme and will provide an operational space weather prediction package to the UK Meteorological Office by 2023. The SWEEP package operates a robust, complimentary framework of multiple models using different boundary conditions including the tomography described in this paper, and both simple and more sophisticated magnetic models \citep{yeates2018,weinzierl2016,gonzi2021}.

Approaches to space weather forecasting can be broadly placed in two groups: simulations that use observations of the Sun’s photosphere to drive a model of the solar wind (e.g., Wang-Sheeley-Arge (WSA)/ENLIL: \citealt{arge2000,WangSheeley1990}) and a persistence based approach which extrapolates the future behavior of the solar wind based on its past behavior over various timescales \citep[][]{owens2018}. A persistence based approach assumes that the ambient solar wind does not evolve drastically over a solar rotation. This is shown in observational data by a weak recurrence in geomagnetic activity and solar wind conditions \citep[][]{Diego2010} over a $\sim$ 27 day period. Hence a persistence approach can provide a baseline for comparison of solar wind forecasting models \citep[][]{owens2013}. The heliospheric simulations are primarily based on remote solar observations and depend on photospheric magnetic field observations to build a model of the corona (e.g, MAS: \citealt{Linker1999}; \citealt{Riley2012}, AWSom: \citealt{van_der_Holst_2014}, etc.). It is the modelled conditions of the outer corona that forms the inner boundary for heliospheric solar wind models (e.g, ENLIL: \citealt{ODSTRCIL2003497}, EUHFORIA: \citealt{Pomell2018}, HUXt: \citealt{Owens2020}). 

The Magnetohydrodynamic Algorithm outside a Sphere (MAS) coronal model uses photospheric magnetic field observational data in order to gain the magnetic field configuration at the solar wind base (\citealp{Linker1999}; \citealp{Riley2012}). MAS derives an inner boundary condition at 30\Rs\ that can be used in solar wind heliospheric models to predict solar wind conditions at 1AU \citep{Owens_Riley_2017}. In this work, the MAS inner boundary condition is used to benchmark the results of the tomography derived inner boundary condition. MAS adopts a simplified coronal `polytropic' model which assumes the thermal pressure is greater or equal to the magnetic pressure at distances closer to the sun (i.e., $\beta$ $\geq$ 1). This is used to empirically convert the pressure and density, which are gained via simplified physical laws, into a solar wind solution \citep{Parker1964}. Magnetohydrodynamic (MHD) equations are used and integrated forward in time until the solar wind parameters reach a steady-state and give a full three dimensional state of the solar wind at heights greater than the solar wind Alfvén point (\citealp{Linker1999}; \citealp{Riley2012}).  MHD heliospheric models use this information at 30\Rs\ as inner boundary conditions in order to propagate the solar wind conditions to 1AU (\citealp{ODSTRCIL2003497}; \citealp{Owens2020}; \citealp{Riley2011}). Three-dimensional MHD models offer a comprehensive physical model of the solar wind at large scales, but are computationally expensive.

\citet{Riley2011} proposed that the magnetic, gravitational, and pressure gradient forces of the solar wind plasma can be neglected at distances greater than 30\Rs, and that a purely radial flow of the ambient solar wind plasma can be assumed, thus vastly reducing the complexity of the MHD equations. Heliospheric models that use this reduced physical approach have a greatly increased computational efficiency at the expense of reduced physics whilst still yielding results comparable to full 3D MHD models \citep[e.g., ][]{Owens2020}. \citet{Owens2020} has adapted this reduced physical model for the time domain, namely the `Time-dependent Heliospheric Upwind Extrapolation' (HUXt) model. The model complexity is reduced further when limited to only radial components of the equatorial plane, allowing the model to run on a standard desktop computer in seconds, even with a moderate angular ($\sim$2.8\de), radial (1.5\Rs), and time ($\sim$ 4 hour) resolution.

Whilst space weather forecasting has developed enormously over the past few decades, there is still large room for further improvement. One of the main constraints on the accuracy of space weather forecasting is the lack of adequate observational data that constrains direct empirical models between the photosphere and Earth. According to the concluding sentences of \citet{macneice2018}: `the pace of [physical model] development has outstripped the pace of improvements in the quality of the input data which they consume, and until this is remedied, these models will not achieve their full forecasting potential'. This statement is a strong argument for new instruments and missions that are focused on operational space weather to provide higher quality data. It is also an argument for full exploitation of current resources: new or improved constraints on coronal structure and density at the coronal-heliosphere boundary are therefore important. Recent efforts are focused on improving the diagnostics from photospheric observations through advanced magnetic modelling, or to use alternative observations such as radio scintillation \citep{gonzi2021}. Our efforts are focused on using coronagraph observations through coronal rotation tomography to provide a direct constraint on solar wind models. A less direct approach developed by \citet{poirier2021} is to use coronagraph observations to constrain magnetic models without resolving the line of sight (LOS).

There are ample, routine, observations made of the coronal-heliospheric boundary region by space-based coronagraphs (e.g., LASCO \citealp{Brueckner1995}, COR2: \citealp{howard2002}). These are largely neglected in the context of ambient solar wind modelling due mainly to the LOS problem: given the complex spatial distribution of high- and low-density streams along the LOS it is impossible, from a single observation, to estimate this distribution. Coronal rotational tomography techniques aim to find a distribution of electron density in a 3D corona which best satisfy a set of coronagraphic polarized brightness (\pB) observations made over half a solar rotation (half a rotation since both east and west limbs are observed), subject to some reasonable assumptions such as the smoothness of the reconstruction, thus helping to rectify the LOS problem. A comprehensive review is given by \citet{aschwanden2011}. An iterative regularised least-square fitting method was presented by \citet{frazin2000}, and developed and applied to other cases \citep[e.g.,][]{butala2005}. A similar method has been applied to very low heights in the corona \citep{kramar2014}, and expanded to include a time-dependency \citep[e.g.,][]{vibert2016}. A novel method for creating qualitative maps of the distribution of coronal structure was introduced by \citet{morgan2009}, resulting in a comprehensive study of coronal structure over a solar cycle \citep{morgan2010structure} and measurements of coronal rotation rates \citep{morgan2011rotation}. Machine learning approaches are currently being developed \citep{jang2021}. A new quantitative method based on spherical harmonics is presented by \citet{Morgan2019}, utilising the calibration and processing methods of \citet{morgan2015}. Further refinements to the method, and initial results, are presented in \citet{morgan2020}, and a study of coronal rotation rates based on the tomography is made by \citet{edwards2021}. The method is based on a spherical harmonic model of the coronal electron density, and gives reconstructions at distances of 4 to 10\Rs. At these heights and above, the corona can be assumed to flow radially outward. The tomography results when compared at different heights, confirm this radial structure \citep{morgan2020}. A desirable goal over the next decade would be an unified approach, where solar wind models are driven by coronal models based on as many empirical constraints as possible, including magnetic field extrapolations, coronal density estimations, and any other routine empirical constraints such as outflow velocity estimations. A crucial advancement would be the inclusion of a time-dependent inner boundary condition based on time-dependent magnetic models \citep[e.g.][]{yeates2008,weinzierl2016}, and time-dependent tomography \citep[e.g.][]{morgan2021,vibert2016}. \citet{morgan2021} state that the time-dependent method requires further development for routine use, thus our current work uses the static tomography method of \citet{morgan2020}.

The aim of this work is to improve the accuracy of predictions of the ambient solar wind velocity at Earth, and to investigate the relationship between coronal electron density and solar wind velocity at a distance of 8\Rs. Although iterative tomography methods have been coupled with MHD models to forecast space weather before \citep[e.g., HELTOMO][]{Jackson2020}, these models and studies are primarily focused on Coronal Mass Ejections (CME) \citep{Jackson_2010} and use an iterative integrated kinematic model in order to predict ambient solar wind conditions \citep{Jackson2013}. Such heliospheric models incorporate CME models such as the cone model \citep{Odstrcil2004}, which use observational constraints on CME characteristics in order to model CME propagation throughout the heliosphere. CMEs are not considered further in this work, although our method and results are relevant to CME propagation and arrival time predictions.

This work proposes to use the tomography results as a new inner boundary condition for heliospheric solar wind models (HUXt), describes an initial implementation, and presents initial results compared to boundary conditions based on MAS as a proof of concept. The methodology used in this study as well as the simple empirical relationship used to derive solar wind velocity from density at a distance of 8\Rs, are described in section \ref{method}. The iterative method used to improve the model's match to \insitu\ data through adjustment of input parameters is given in section \ref{Vsolution}. MAS inner boundary conditions, when coupled with 3D MHD heliospheric models, have produced results that compare well with \insitu\ (OMNI) bulk solar wind velocity data \citep[e.g.,][]{Owens_Riley_2017}. The optimised model output is compared to results gained by using MAS-derived inner boundary condition and \insitu\ data obtained via Operating Missions as Nodes on the Internet (OMNI) satellite network in section \ref{Section:Validation of model}. The tomography-based model is then applied to dates at different stages of solar cycle 24 in section \ref{Section:application to other dates}. Section \ref{persistence} explores the operational feasibility of the tomography inner boundary condition by adopting a persistence-based approach. Section \ref{Ensemble} presents an implementation of an ensemble model that demonstrates how uncertainties can be quantified by the system. Conclusions are given in section \ref{conclusions}.
}

\section{Method}
\label{method}
This section gives a brief overview of the tomography method (section \ref{section:TomMaps}), outlines the method to generate the inner boundary velocity condition from the tomography density maps (section \ref{section:Den-Vel model}), gives an overview of the the heliospheric solar wind model (section \ref{section:HUXt}), and an overview of the Dynamic Time Warping (DTW) algorithm and how this will be exploited in the context of this study (section \ref{section:DTW}).

\subsection{Tomography Maps of the Solar Corona}\label{section:TomMaps}

\begin{figure}[h!]
    \centering
    \includegraphics[width=0.85\textwidth]{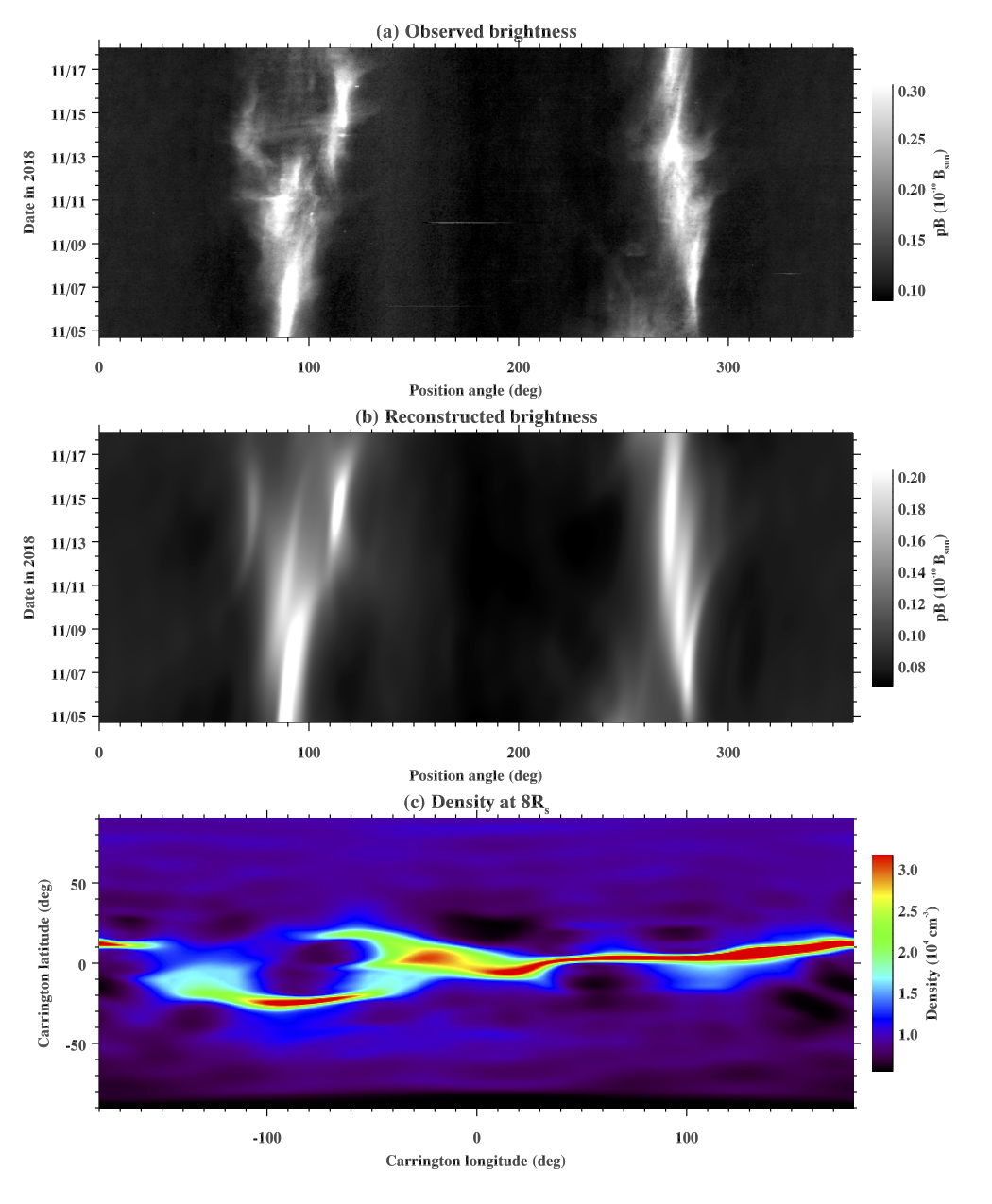}
    \caption{(a) The observed coronal polarized brightness at a distance of 8\Rs, plotted as a function of position angle (measured counter-clockwise from solar north), and time. This is used as input to the tomography method. (b) The reconstructed polarized brightness, or synthetic observations gained from the tomographical density distribution. (c) The electron density at a distance of 8\Rs, mapped in Carrington longitude and latitude. }
    \label{Tomo_map_CR2210}
\end{figure}

The COR2 coronagraphs are part of the Sun Earth Connection Coronal and Heliospheric Investigation (SECCHI:  \citealp{howard2002}) suite of instruments aboard the twin Solar Terrestrial Relations Observatory (STEREO A \& B: \citealp{kaiser2005}). Density maps are calculated from COR2A data in 3 main steps:
\begin{itemize}
\item Calibration is applied to COR2 polarized brightness (\pB) observations over a period of half a Carrington rotation ($\pm$1 week from the required date) using the procedures of \citet{morgan2015}. The processing includes a method to reduce the signal from coronal mass ejections (CMEs) \citep{morgan2012cme}. Following calibration, the data is remapped in solar polar co-ordinates, and an annular slice at the required distance from sun center extracted over several hundred observations taken over the time period. Figure \ref{Tomo_map_CR2210}a shows an example of data used as input for the tomography. For this example, the central date is 2018/11/11 12:00, the data spans a period from 2018/11/04 00:00 to 2018/11/17 23:08, and the distance chosen is 8\Rs. This date range corresponds approximately to the mid-date of Carrington rotation (CR) 2210.
\item Tomography is applied to the data using the regularized spherical-harmonic optimization approach of \citet{Morgan2019}. The method is based on a spherical harmonic distribution of density at the height of reconstruction, with a $r^{-2.2}$ decrease in density above this inner height, thus a purely radial density structure is prescribed. Line-of-sight integrations are made of the spherical-harmonic based densities, thus a set of brightnesses are gained, one for each order, with the lines-of-sight corresponding to the COR2A observations. The problem is then reduced to finding the coefficients of each order based on a regularised least-squares fitting between the observed brightness and the spherical harmonic brightnesses. In this work, a $22^{th}$ order spherical harmonic basis results in a density reconstruction with 540 longitude and 270 latitude bins for a chosen distance (restricted to between 4 and \app10\Rs, with this range limited by the instrument's useful field of view). In this work, we use only the reconstructions for a distance of 8\Rs. Note also that the tomography reconstruction is static - it has no time dependence.
\item High-density streamers are narrowed, and a correction for `excess' density, possibly F-corona contamination \citep{morgan2007fcorona}, is applied according to the method of \citet{morgan2020}. The narrowing method is applied based on the gradients within the initial tomography density map, with the degree of narrowing controlled by a single parameter. This parameter is adjusted to find an optimal fitting to the observations. The `excess density' is estimated based on an analysis of densities within large low-density regions (coronal holes) between tomography maps made at a range of distances (4 to 8\Rs), and a consideration of mass flux. 
\end{itemize}

Figure \ref{Tomo_map_CR2210}a shows the observed coronal polarized brightness. This distribution is very similar to the reconstructed polarized brightness shown in figure \ref{Tomo_map_CR2210}b. The reconstructed brightness is the result of synthetic observations that use the electron density distribution. The tomography density map resulting from the above steps, is shown in figure \ref{Tomo_map_CR2210}c. This visualises the distribution of electron density at a distance of 8\Rs\ resulting from the steps described above. Note that this is a static reconstruction, since it gives a non-time-dependent density distribution that is spatially smooth and best satisfies the data.

The distribution of the reconstruction shows a good agreement with the observed, and all the large-scale streamer features are well reconstructed. However, the reconstruction lacks the fine-scale detail of the observed, including a general `fuzziness' of the brightness and temporal changes over timescales of less than a day. To reconstruct fine-scale detail, it is necessary to apply the time-dependent tomography described by \citet{morgan2021}. There is considerable small-scale variation in the nascent slow solar wind \citep[e.g.,][and references within]{alzate2021}, and the evolution of this variation in the heliospheric wind is an active field of research. For operational forecasting of the ambient solar wind, our immediate problem is to model the large spatial scale and longer-timescale variations, and the static tomography reconstruction is adequate for this purpose.

\subsection{Generating the Inner Boundary Condition} \label{section:Den-Vel model}

\begin{figure}[h!]
    \centering
    \includegraphics[width=0.9\textwidth]{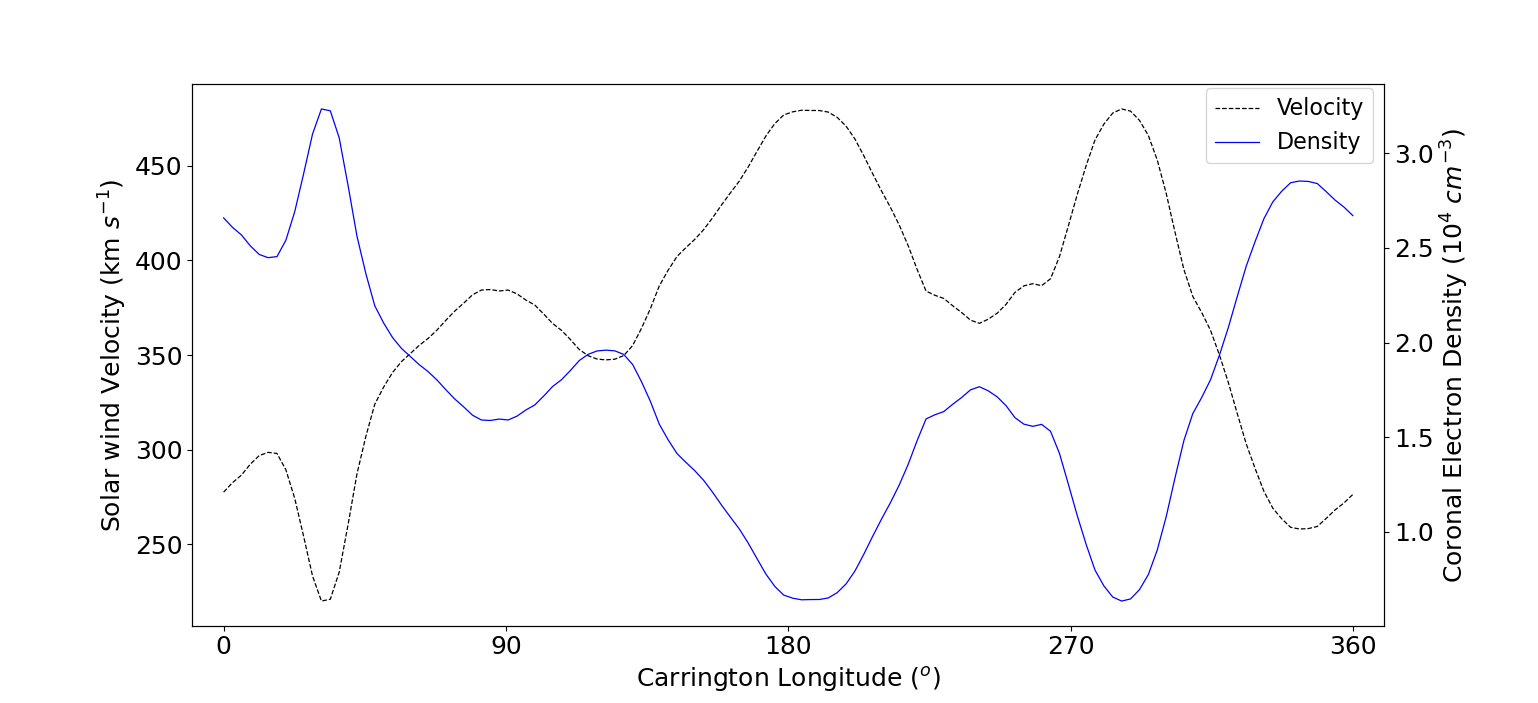}
    \caption{Coronal electron density (blue, right axis) taken from the tomographical map of figure \ref{Tomo_map_CR2210}c at Earth's latitude, and solar wind velocity at 8\Rs (black, left axis) gained from the density using the simple inverse relationship of equation \ref{P_V_Empricial_Model}. For this example, $V_{max} = 480$\kms\ and $V_{min} = 220$\kms.}
    \label{Den-Vel}
\end{figure}

We wish to use the tomography density map as an inner boundary condition to the HUXt model, and to compare the resulting solar wind velocities near Earth with the results of using a MAS-based inner boundary. HUXt allows the user to set the heliocentric distance of the inner boundary. In this work, a distance of 8\Rs\ is used for the tomography-based inner boundary, and 30\Rs\ for the MAS-based inner boundary. For the purpose of solar wind modelling in the Sun-Earth equatorial plane, a slice of density is extracted from the tomography map at Earth's Carrington latitude at the initiation of the Carrington rotation (this latitude is 4.8\de\ for the 2018 November example). The heliospheric model requires only 128 solar wind velocity values at equally spaced longitudes with increments of\app2.8$^{\circ}$. The extracted tomography longitudinal profile is rebinned to this size. Figure \ref{Den-Vel} shows the density profile at this latitude as a blue curve.

The solar wind model requires a set of radial outward velocities as an inner boundary condition. We adopt a simple linear inverse relationship between densities and velocities, approximately consistent with the general properties of the solar wind as revealed by both \insitu\ and remote observations \citep[e.g.,][]{Allen2020, habbal1997, Schwenn2006}, given by:

\begin{equation} \label{P_V_Empricial_Model}
    V = V_{max} - \left[ \left(\frac{n- n_{min}} {n_{max} - n_{min}} \right) \left( V_{max} - V_{min}\right)             \right], 
\end{equation}

where $V$ and $n$ are the solar wind velocity and electron particle density respectively, $n_{max}$ and $n_{min}$ are the maximum and minimum densities in the equatorial plane respectively, and $V_{max}$ and $V_{min}$ are model parameters specifying the maximum and minimum solar wind velocity at the height of the inner boundary (8\Rs). Whereas the density values ($n_{max}$ and $n_{min}$) are defined by the tomography data at 8\Rs, the range of the velocity values is unknown. This highlights the need for an optimisation process with the aim of finding optimal values of $V_{max}$ and $V_{min}$. This optimisation process is described in section \ref{Vsolution}. Figure \ref{Den-Vel} demonstrates the conversion of the electron density to the solar wind velocity at the inner boundary for the 2018 November (CR2210) example with $V_{max}$ and $V_{min}$ set to their optimal values of 480 and 220\kms\ respectively. Note that these optimal values are found in a following section.

This simple inverse linear relationship of equation \ref{P_V_Empricial_Model} is likely oversimplistic compared to the true relationship between density and velocity in the corona, but serves the purpose of providing a proof of concept of a direct relationship for this study. Future efforts will experiment with optimising this empirical relationship, for example, an inverse Sigmoid function or an exponential relationship may better model the likely bimodal slow and fast wind patterns in the nascent solar wind, and may reveal a global model that can provide an optimised model of the solar wind over several years, or a solar cycle. We note that converting the density into an estimate of velocity is a similar approach to that used by magnetic models, where an empirical relationship is required to transform the magnetic field distribution to velocity \citep[e.g.][]{gonzi2021}.

\subsection{Heliospheric Upwind Extrapolation Model} \label{section:HUXt}

HUXt is an incompressible solar wind model that solves a reduced Burgess equation along 1D vectors of velocities in the radial domain. The `upwind' conditions only allow outward flow and thus forbids any sunward motion of the solar wind. Therefore, considerations for stream interaction are included through adjustments in solar wind speed to uphold the upwind conditions (or at a greater distance in the radial co-ordinate) \citep{Riley2011}.  For the purposes of forecasting, the results of this model compare well to full 3D MHD models for both predictions of the ambient solar wind and CME events (\citealp{hinterreiter2021}; \citealp{Owens2020}). In this work we use the static tomographical reconstruction and therefore a time-constant boundary condition. 

The HUXt model includes a parameter to account for residual solar wind acceleration. \citet{Riley2011} used an exponential function to impose a velocity profile that approached its final value asymptotically over a distance range between the lower boundary and \app50\Rs\ (an acceleration parameter of 0.15 which is built in to the HUXt model). In this work, we use this acceleration parameter for both slow and fast wind streams. Investigating this parameter is a central focus of our future work: it is well known that the slow wind reaches its final velocity at a greater distance from the sun compared with that of the fast \citep[e.g.,][]{Schwenn1990}.

HUXt uses a five day `spin-up' time. All longitudinal and radial model points are initialised with a value of 400\kms. The velocity at the inner boundary condition is rotated through\app-66\de\ (equal to the solar rotation over 5 days), and the solar wind conditions are iteratively propagated outward whilst the boundary condition is rotated forward with time. Therefore, after the five day `spin-up' period, propagated solar wind velocity values have reached distances beyond the orbit of Earth. Following the spin-up period the solar wind velocity conditions are simulated forward in time by 27.27 days (or one rotation). Results that are generated during the spin-up period are discarded from the model output \citep{Owens2020}. For this study, the inner boundary condition is used as a static state which is not altered during the model run.  

The evolution of solar wind velocity can be plotted as a function of time at any point in space within the model. In this work, we limit the results to Earth, allowing a direct comparison with \insitu\ measurements. We use the reduced 5-min resolution combined satellite network data provided by OMNI. Sporadic short periods of missing solar wind velocity values are linearly interpolated and then binned to form an hourly average with an associated standard deviation. This is then smoothed with a 10-hour moving window average.  

\subsection{Dynamic Time Warping}\label{section:DTW}
Dynamic Time Warping (DTW) is as an effective algorithm to quantify the agreement between two time series and is used here to compare the model and \insitu\ solar wind velocities. The DTW algorithm was initially used to aid automated speech recognition and has recently been used in a variety of fields such as economics \citep{Franses2020}, biology \citep{Skutkova2013}, and space weather (\citealp{Samara2021}; \citealp{Owens_2021}).

DTW requires two vectors (1D arrays, $A$ and $B$) as input. Vectors $A$ and $B$ are not required to be the same length. The Euclidean distance is calculated between every point in set A to every point in set B and is thus assigned a cost function for every possible alignment between the two data sets. This DTW cost or `DTW distance' metric is then minimised to find the optimal alignment between data set A and B \citep{Berndt1994}. Effectively, DTW is non-linearly stretching and compressing each data set by connecting like for like structures between the sets. More efficient versions of this algorithm have been generated for reduced computational expense such as the Python FastDTW algorithm package used in this study \citep[See:][]{Salvador2004}.

The requirements of the DTW algorithm is as follows: 
\begin{itemize}
   \item The two sets are ordered in time.
   \item Both sets start and end at the same time, or $A_0$ is anchored to $B_0$, and $A_{n-1}$ is anchored to $B_{m-1}$, where $n$ and $m$ are the number of elements in $A$ and $B$ respectively. 
   \item Every element in data set $A$ will be matched with at least one corresponding data point from data set $B$, and vice versa. 
   \item The optimum path must not `cross' between elements (for example, if $A_{i}$ is paired to $B_{j}$, then $A_{i+1}$ or later cannot be paired with $B_{j-1}$ or earlier).
 \end{itemize}

A metric used in this study in order to quantify the DTW distance is the Sequence Similarity Factor (SSF). SSF, as defined by \citet{Samara2021}, is described in eq \ref{eq:SSF}: 

\begin{equation} \label{eq:SSF}
    SSF=\frac{DTW_{Dist}(O,M)}{DTW_{Dist}(O,\Bar{O})}
\end{equation}

Where $M$ represents the modelled solar wind velocity and $\Bar{O}$ represents the average magnitude of the observed \insitu\ data ($O$). SSF allows a direct comparison of the modelled data to that of the averaged \insitu\ data and provides a surrogate score of the model. For context, if SSF is \textgreater\ 1, the predicted solar wind velocities are worse than that of a constant mean observed solar wind value across the full time period of the prediction. If SSF = 0, a perfect prediction has been made and the modelled data matches the \insitu\ data exactly.

\section{Results}
\subsection{Velocity Optimisation using Dynamic Time Warping}\label{Vsolution}
This section describes and implements a simple method to derive optimised values of both the $V_{min}$ and $V_{max}$ terms, and shows how this approach yields a far improved agreement between the model and measured solar wind velocities at Earth, as well as time of arrival of fast solar wind streams.

\begin{figure*}[h!]
        \centering
        \subcaptionbox{\label{Overestimated_CR2210}}[0.495\textwidth]{\includegraphics[trim={0cm 0 0 0cm},clip,width=0.495\textwidth]{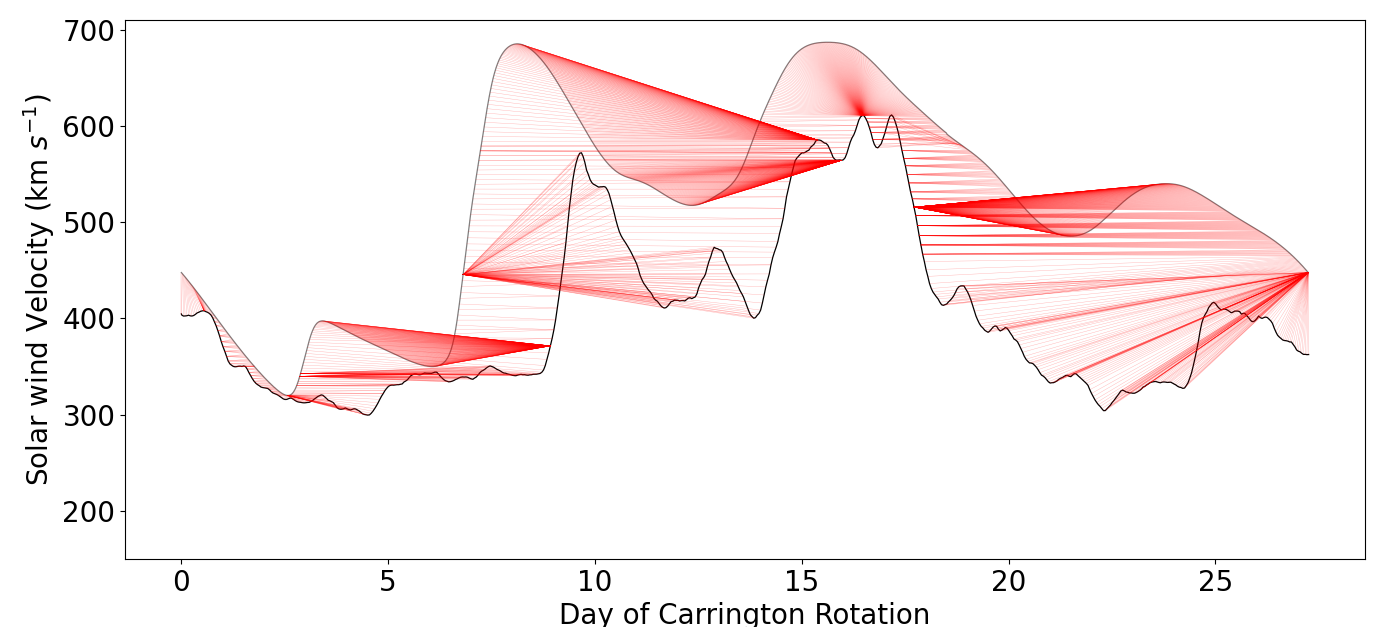}}%
        \subcaptionbox{\label{underestimated_CR2210}}[0.495\textwidth]{\includegraphics[trim={0cm 0 0 0cm},clip,width=0.495\textwidth]{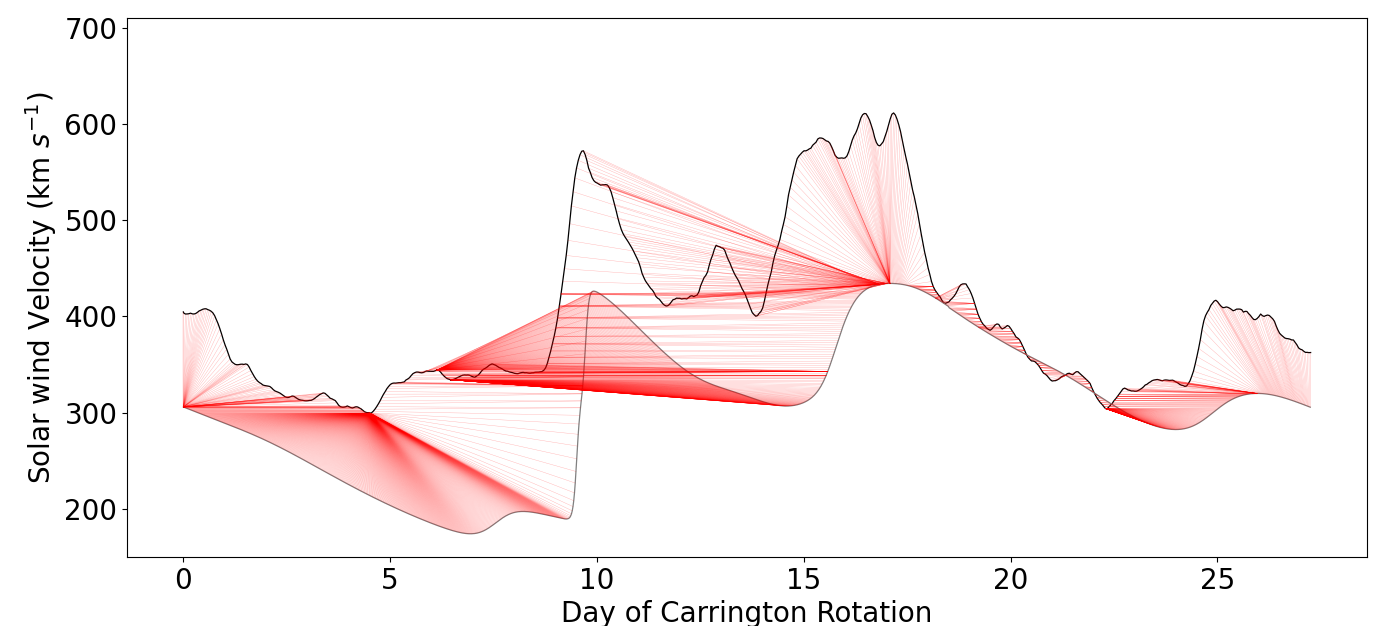}}
        
        \caption[ Over_under estimation  ]
        {DTW optimal path (red) between \insitu\ data and two Tomography/HUXt model runs for CR2210 with a) overestimated $V_{max}$ and $V_{min}$ terms of 600 and 250\kms\ respectively, and b) underestimated $V_{max}$ and $V_{min}$ terms of 380 and 110\kms\ respectively.} 
        \label{CR2210_overestimated_underestimated}
    \end{figure*}

Figure \ref{CR2210_overestimated_underestimated} demonstrates the effect of using arbitrary (and inaccurate) $V_{max}$ and $V_{min}$ velocity terms on the solar wind model values at Earth. $V_{max}$ and $V_{min}$ parameters are set at 600\kms and 250\kms (shown in figure \ref{Overestimated_CR2210}) and 380\kms and 110\kms (shown in figure \ref{underestimated_CR2210}). This figure also visualises the DTW connections between the model and OMNI measurements along the optimum DTW path (shown in red) with the overall DTW distance being visualised by the sum of the length of the red lines. Both overestimation and underestimation of the velocity parameters will cause inaccurate predictions of the solar wind at Earth and thus give a larger DTW path distance compared to a model run with optimised or `best fit' velocity parameters. For example, the magnitude of the DTW path distance metrics for the model runs seen in figure \ref{CR2210_overestimated_underestimated} a) and \ref{CR2210_overestimated_underestimated} b) are $6.03\times 10^4$ (SSF value of 1.34) and $7.66 \times 10^4$ (SSF value of 1.71) respectively. Therefore, both model runs shown in figure \ref{CR2210_overestimated_underestimated} provide a greater DTW distance compared to a constant average observed velocity across the full Carrington rotation. In order to obtain velocity values that give an optimal fit (or lowest DTW distance) between the model and \insitu\ data, the efficiency of the HUXt model is exploited in an exhaustive search method. The tomography/HUXt model is run repeatedly with incrementally changing  $V_{max}$ and $V_{min}$ terms, and the total DTW distance is recorded for each run.

\begin{figure*}[h!]
        \centering
        \subcaptionbox{\label{DTW_V_sol_2210}}[0.495\textwidth]{\includegraphics[trim={0cm 0 0 0cm},clip,width=0.495\textwidth]{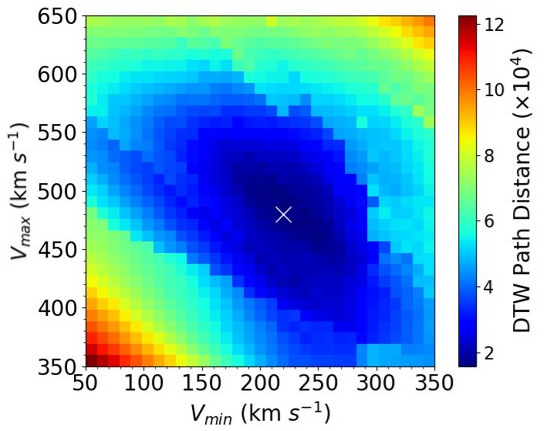}}%
        \subcaptionbox{\label{DTW_V_sol_2210_MAS}}[0.495\textwidth]{\includegraphics[trim={0cm 0 0 0cm},clip,width=0.495\textwidth]{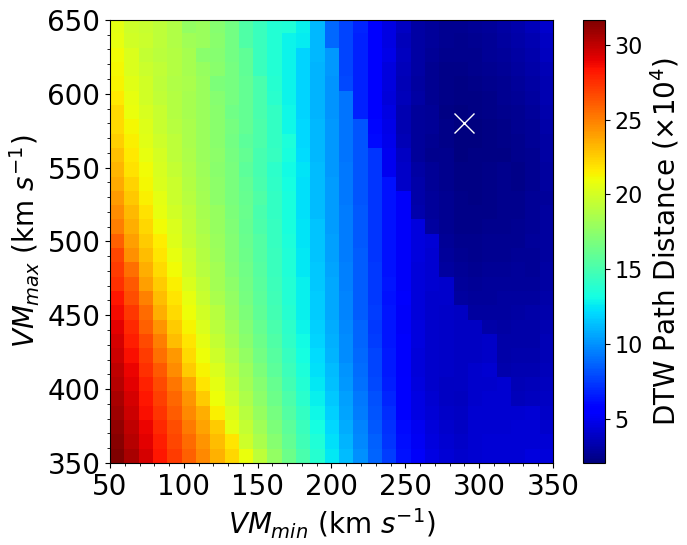}}
        
        \caption[ Over_under estimation  ]
        {(a) Contour plot of DTW path lengths with varying magnitude of the $V_{max}$ and $V_{min}$ terms for the tomography derived inner boundary condition. The minimum DTW path distance is marked by a white cross, and this point gives the optimal values of $V_{max}$ and $V_{min}$ of 480 and 220\kms\ respectively. (b) DTW path lengths as a function of varying scale parameters for the MAS-derived inner boundary condition with the white cross showing the optimum $VM_{min}$ and $VM_{max}$ parameters of 280 and 580\kms\ respectively.} 
        \label{CR2210_Heat_maps}
    \end{figure*}

\begin{figure}[h!]
        \centering
        
        \subcaptionbox{\label{DTW temporal path data}}[\textwidth]{\includegraphics[trim={0 0 0 0cm},clip,width=\textwidth]{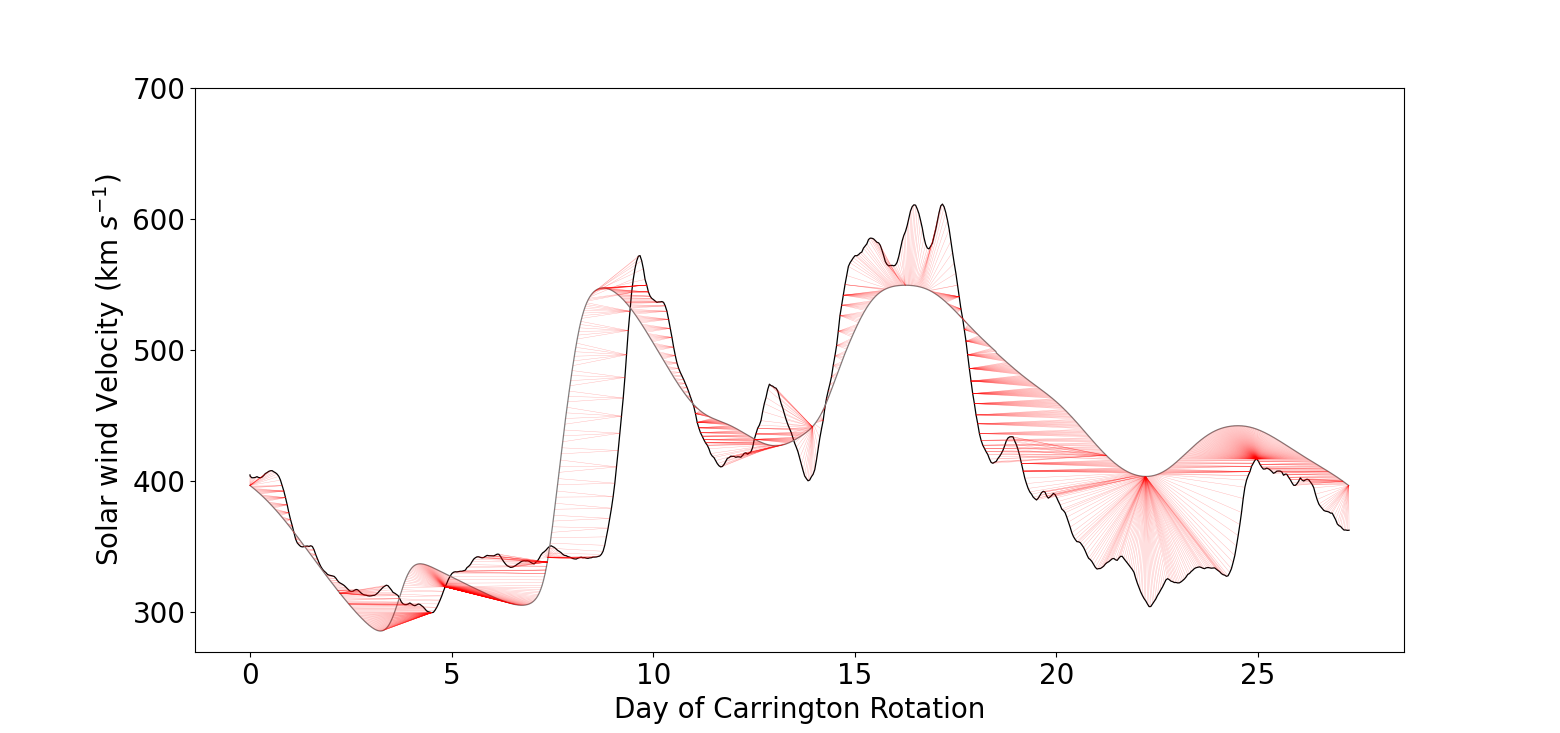}}%
        \vskip\baselineskip
        \subcaptionbox{\label{DTW temporal path}}[0.315\textwidth]{\includegraphics[trim={0 0 0 0cm},clip,width=0.315\textwidth]{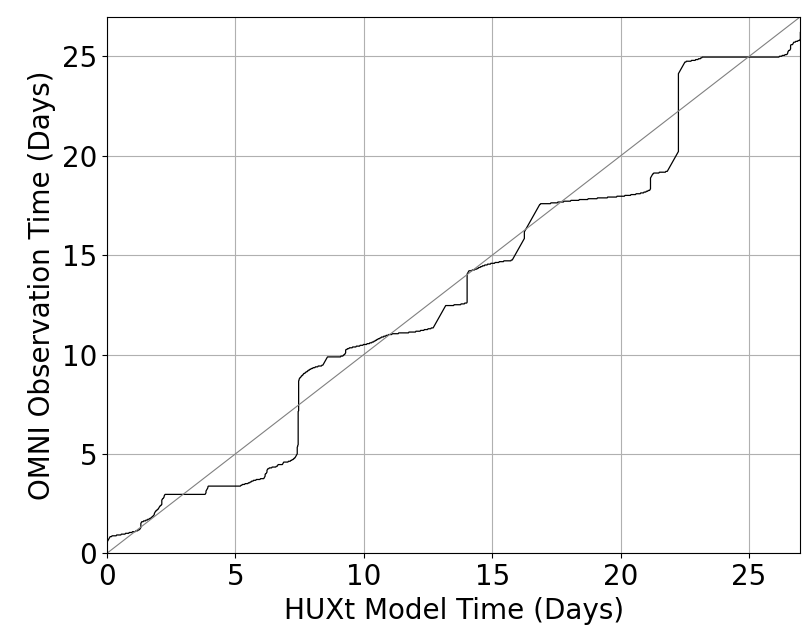}}
        \subcaptionbox{\label{T_diff_2210}}[0.36\textwidth]{\includegraphics[trim={0 0 0 0cm},clip,width=0.35\textwidth]{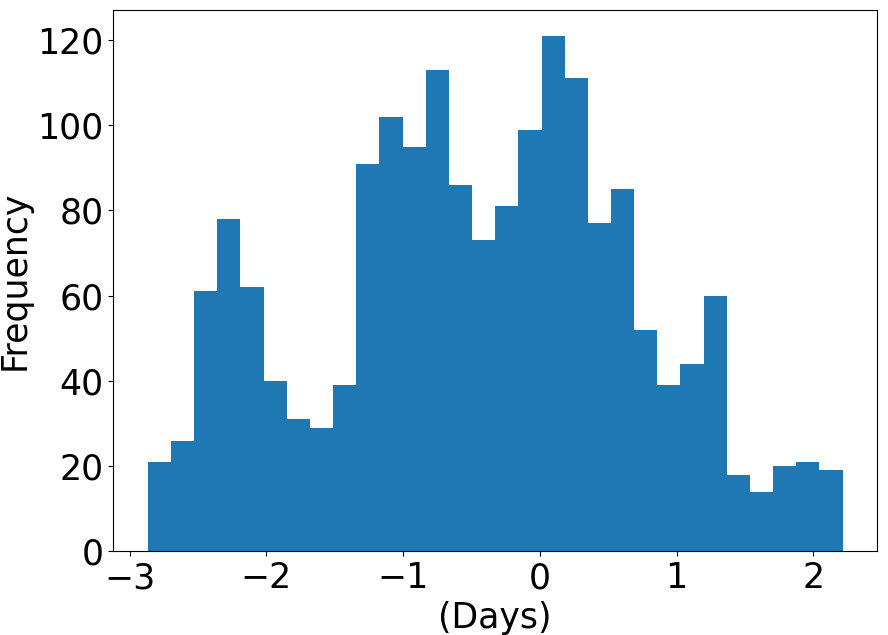}}
        \subcaptionbox{\label{V_diff_2210}}[0.3\textwidth]{\includegraphics[trim={0 0 0 0cm},clip,width=0.34\textwidth]{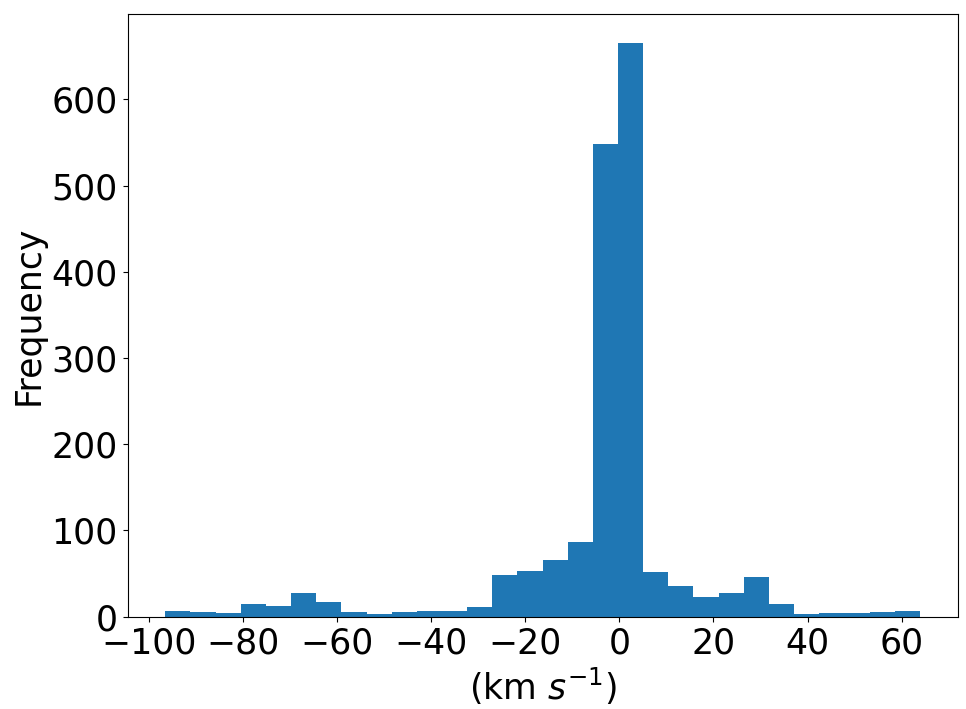}}
        
        \caption[ The average and standard deviation of critical parameters ]
        {Further analysis of the optimal DTW path. (a) Comparison of modelled solar wind velocities at Earth (grey) with \insitu\ data (black) and DTW optimised path (red) (b) Alignment of the modelled and \insitu\ data along the optimised DTW path with respect to time. (c) Difference in time between observation and modelled data of aligned points along optimal DTW path ($\Delta T_{(in situ-model)}$) (d) Velocity difference between observation and modelled data of aligned points along optimal DTW path ($\Delta V_{(in situ-model)}$).} 
        \label{2210_path}
    \end{figure}

Figure \ref{DTW_V_sol_2210} shows the relationship of optimal DTW path distance between the \insitu\ and Tomography/HUXt model solar wind velocity at 1AU with incrementally changing velocity terms in equation \ref{P_V_Empricial_Model}. The velocity ranges used in this study are 50 - 350\kms\ and 350 - 650\kms\ for $V_{min}$ and $V_{max}$ respectively in order to account for a wide velocity range for each parameter while also constraining each parameter to values that are consistent with physical solar wind velocities. We use 30 increments between these extremes (so $30 \times 30=900$ model evaluations). Figure \ref{DTW_V_sol_2210} shows a minimum optimal DTW path distance for the tomography derived inner boundary condition corresponding to velocity magnitudes of 220 and 480\kms\ for $V_{min}$ and $V_{max}$ respectively. This is represented by the white cross in figure \ref{DTW_V_sol_2210}. These values are lower than observed at Earth due to the HUXt heliospheric model incorporating an acceleration parameter. A comparison between smoothed OMNI satellite data and the HUXt output with the optimal $V_{min}$ and $V_{max}$ is shown in figure \ref{DTW temporal path data}, and demonstrates a very strong correlation between the OMNI data and the output of the tomography/HUXt model (for statistical analysis see Table \ref{Table:Calibrated_stats}). The time of arrival of the faster solar wind streams agree to $\pm$1 day with the velocities agreeing with $\pm$20\kms. 

In order for a fair comparison between a tomography and MAS based inner boundary condition (see section \ref{Section:Validation of model}), the MAS inner boundary condition requires a similar exhaustive optimisation process to be applied. This optimisation process effectively scales the initial MAS inner boundary condition between two values ($VM_{min}$ and $VM_{max}$), with the aim of minimising the DTW path distance between the modelled and \insitu\ data. During the optimisation process and for following comparisons, the MAS inner boundary height was set at 30\Rs. The results of this optimisation can be seen in figure \ref{DTW_V_sol_2210_MAS}, with the minimum DTW path distance corresponding to $VM_{min}$ and $VM_{max}$ values of 290 and 580\kms\ respectively.

Further details of the agreement of model and observation in the context of a DTW analysis are shown in figure \ref{2210_path}. Figure \ref{DTW temporal path data} visualises the DTW optimal alignment via the red lines. Figure \ref{DTW temporal path} shows that the DTW path of the optimal alignment rarely deviates over two days from the `ideal' path which represents a near perfect agreement between model and \insitu\ data shown in grey. The largest disagreement is seen at 18-24 days, which is seen in both the longer red lines between data points in figure \ref{DTW temporal path data} and by the biggest dispersion between the DTW optimum path and the ideal path seen in figure \ref{DTW temporal path}. This is due to a model overestimation of speed during this period. The histograms demonstrate the differences in time of arrival and the magnitude of solar wind velocity between the modelled and observational data along the optimum path. Figure \ref{T_diff_2210} shows that there is a bias towards negative values. This suggests the model is predicting a later time of arrival with a mean $\Delta T_{(in situ-model)}$ of -0.45 days, and a standard deviation of 1.16 days.  Figure \ref{V_diff_2210} is also biased towards negative values, suggesting velocity overestimation by the model. The mean $\Delta V_{(in situ-model)}$ is -3.64\kms\ with a standard deviation of 19.79\kms.

\subsection{Validation of Model Output}\label{Section:Validation of model}
\begin{figure}[h!]
    \centering
    \includegraphics[width=\textwidth]{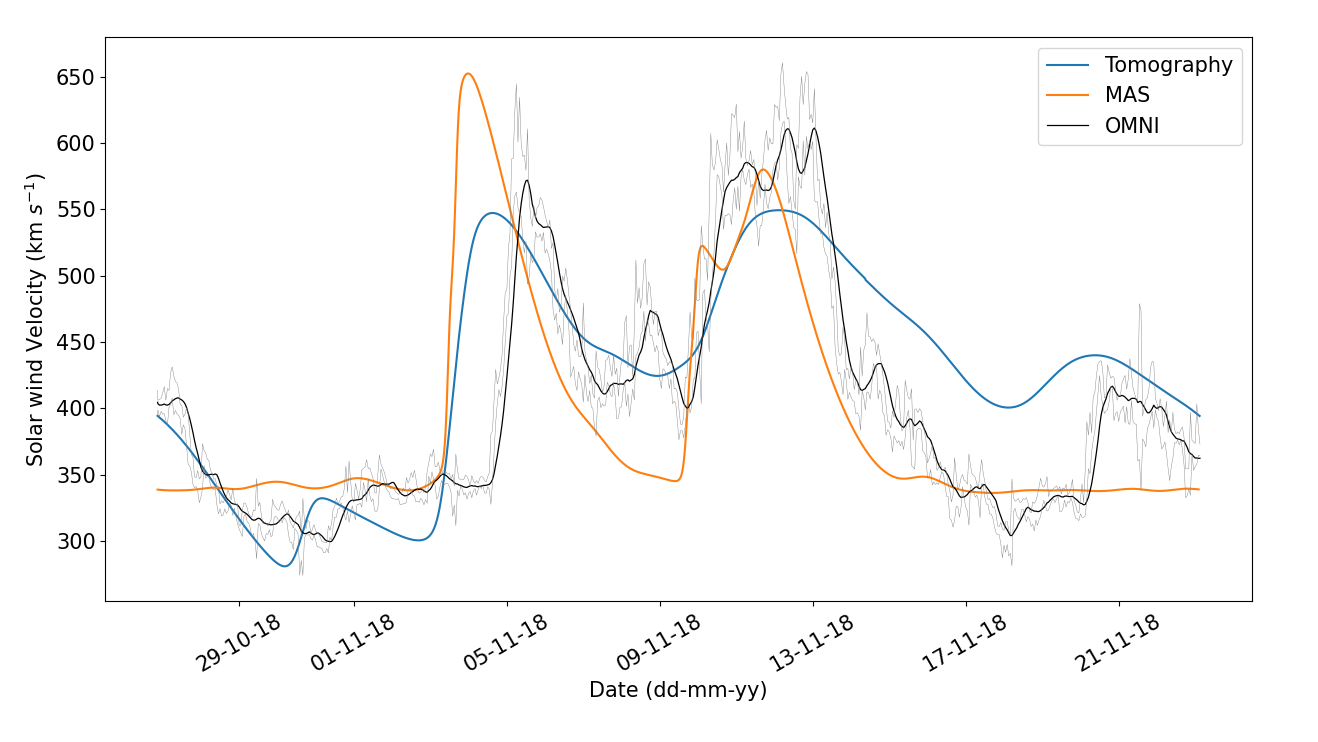}
    \caption{Comparison of optimised tomography/HUXt (blue) and optimised MAS/HUXt (orange) model predictions of the solar wind velocity at Earth with $\insitu$ data (Black) for CR2210.}
    \label{MAS-Tom2}
\end{figure}

Figure \ref{MAS-Tom2} shows a comparison of the CR2210 tomography/HUXt velocity output (Blue) with OMNI \insitu\ measurements (Black) at 1AU and MAS/HUXt velocity output (Orange). The tomography based inner boundary parameters $V_{min}$ and $V_{max}$ set at 220 and 480\kms\ respectively. The MAS inner boundary condition has scale parameters of $VM_{min}$ and $VM_{max}$ of 290 and 580 \kms\ as described in section \ref{Vsolution}.

From figure \ref{MAS-Tom2}, both models have similar profiles, and the main changes between slow and fast wind tend to agree. The first small peak (shown in the tomography data 2018 October 29-31) is not present in the MAS data. The solar wind velocity between the second (2018 November 2-5) and third peak (2018 November 10-14) drops to intermediate velocities ( \app440\kms ) for the tomography-driven model, whilst the MAS model drops to slower velocity (\app350\kms\ ). This is significant as the \insitu\ data, as shown in black in figure \ref{MAS-Tom2}, shows an intermediate velocity (\app440\kms) of the more undisturbed solar wind in between the two fast peaks (2018 November 7-10). The magnitude of solar wind velocity during this time is better represented by the tomography inner boundary condition model.

Table \ref{Table:Calibrated_stats} presents statistical details of the comparison between models. The Pearson correlation coefficient between \insitu\ measurements and models is considerably higher for the tomography/HUXt model. The mean absolute error (MAE) of velocities between measurement and MAS/HUXt is approximately 11\% higher than that for the tomography/HUXt, while also showing a higher SSF. These results show that the general profile and magnitudes of solar wind velocity is closer to the data using the tomography boundary condition compared to the optimised MAS model. This demonstrates both the potential value of tomographical density maps as an inner boundary condition, and the benefit of searching for an optimised velocity range at the inner boundary. Note that such an optimisation would not be possible without an efficient solar wind model such as HUXt.

\begin{table}[h!]
\begin{center}
\begin{tabular}{||c c c c c ||} 
 \hline
 Model & MAE  & Pearson & DTW Path Dist & SSF  \\
 & (km s$^{-1}$) & & $(\times10^4)$ & \\

 \hline\hline
  MAS/HUXt & 53.58 & 0.567 & 2.11 & 0.47\\
 \hline
  Tomography/HUXt & 47.90 & 0.7413 & 1.80 & 0.40 \\ 
 \hline
\end{tabular}
\caption{Statistical analysis of the comparison between the tomography/HUXt model and the MAS/HUXt model, with \insitu\ data.}
\label{Table:Calibrated_stats}
\end{center}
\end{table}

Both models fail to give the exact time of arrival of various features. For example, the rapid rise from slow to fast wind observed on November 4 arrives approximately one day early in both models. The start of the decrease from fast to slow wind on November 13 comes late in the tomography/HUXt model, and the decrease is less rapid than the observed. The main reasons for these differences are listed and discussed in section \ref{conclusions}. Despite the differences, the comparison of tomography/HUXt to \insitu\ data is promising - the large-scale features of the measured velocity are present in the predicted velocities, and the timings are reasonable given the initial simple implementation of the inner boundary condition. The model fails to replicate smaller-scale structures on timescales of a day or less. One reason for this are that the tomography densities are inherently smooth - this is an unavoidable result of finding a static tomographical solution which is discussed further in section \ref{conclusions}. Other reasons for differences between model and measurements include the reduced physics approach of HUXt and the limited resolution of computational modelling.

\subsection{Application to other dates}\label{Section:application to other dates}
\begin{figure}[h!]
    \centering
    \includegraphics[width=0.8\textwidth]{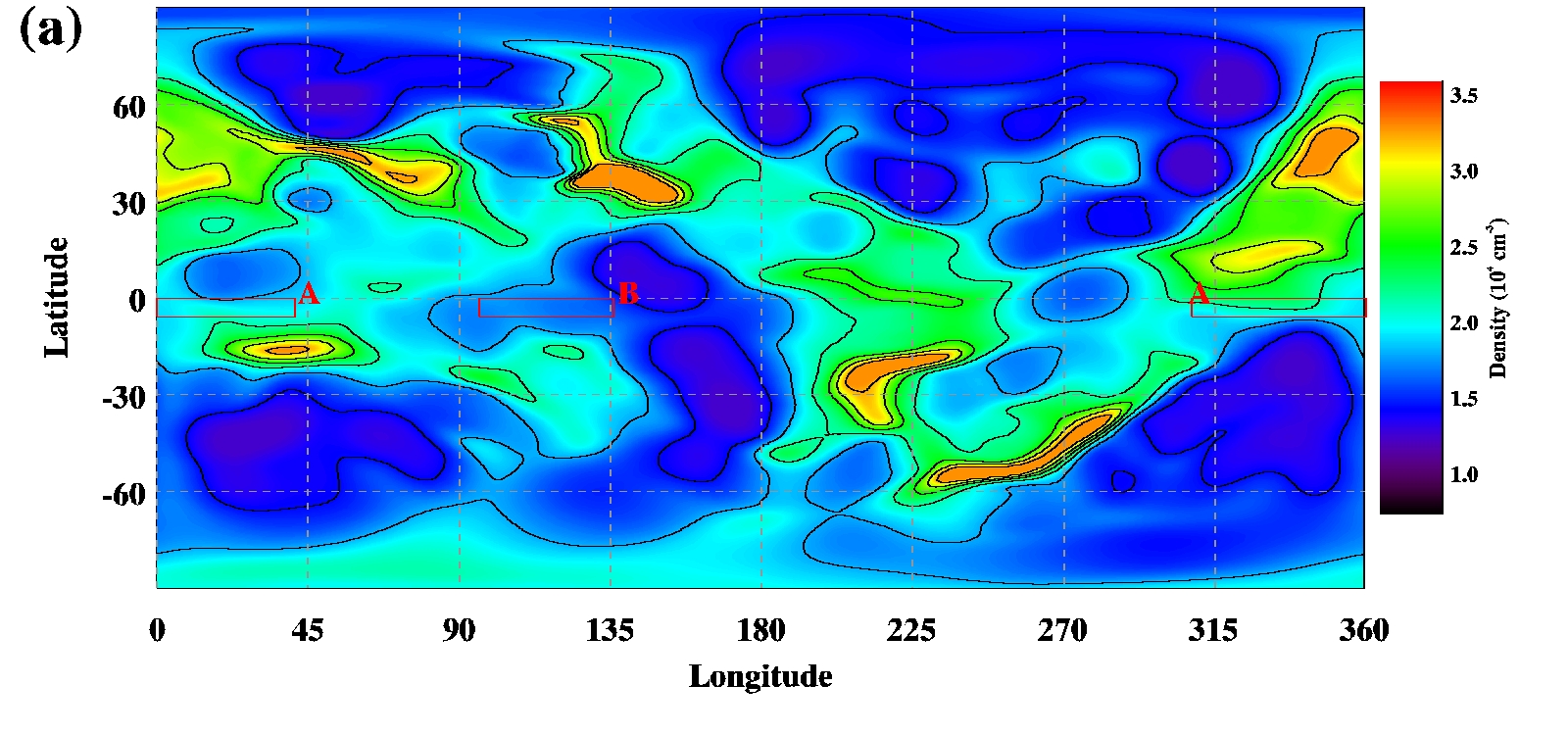}
    \includegraphics[width=0.8\textwidth]{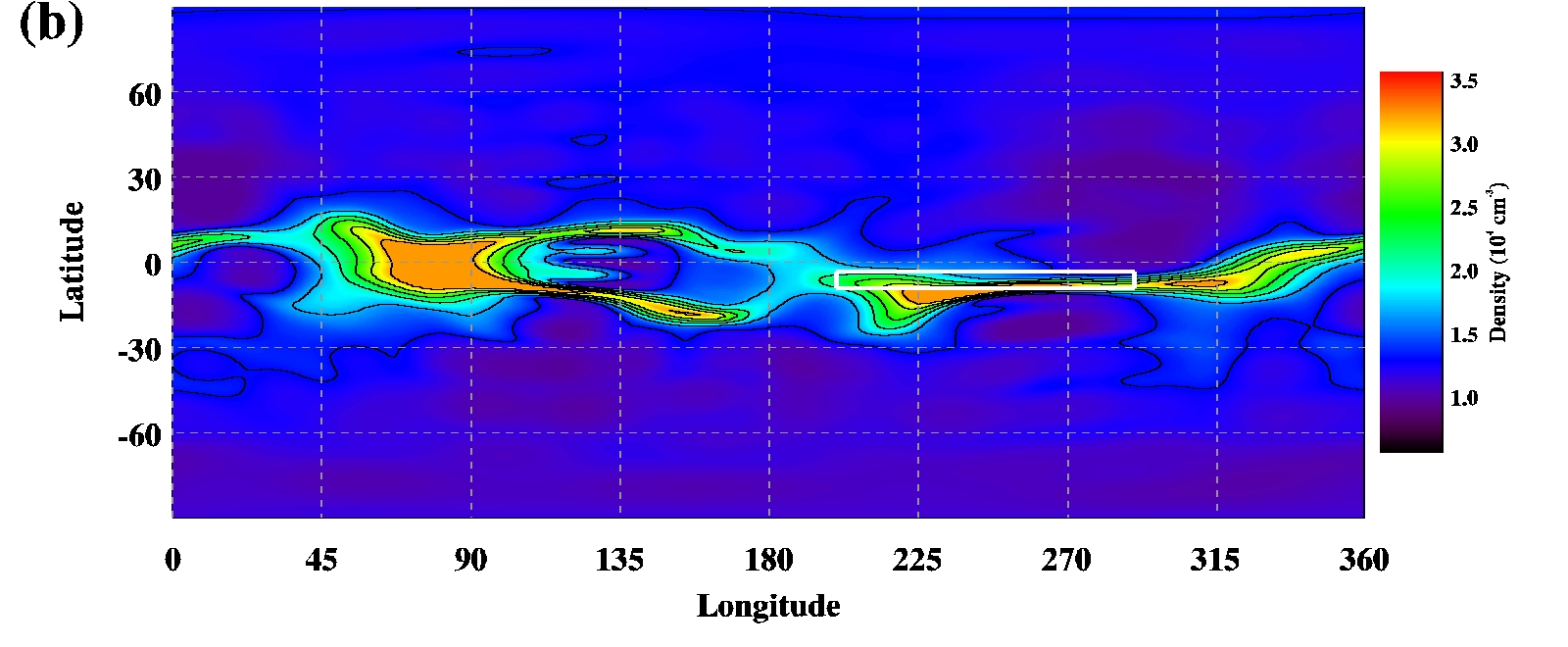}
    \caption{Electron density as estimated using coronal tomography at a distance of 8\Rs, for Carrington rotations (a) 2150.5, and (b) 2202.5. The red boxes labelled A and B in (a), and the white box in (b), are relevant for the ensemble results of section \ref{Ensemble}.}
    \label{tomoresults}
\end{figure}

\begin{figure*}[h!]
        \centering
        
        \subcaptionbox{ CR2202 \label{Best_Fit_CR2202}}[0.595\textwidth]{\includegraphics[trim={0 0 0 0cm},clip,width=0.595\textwidth]{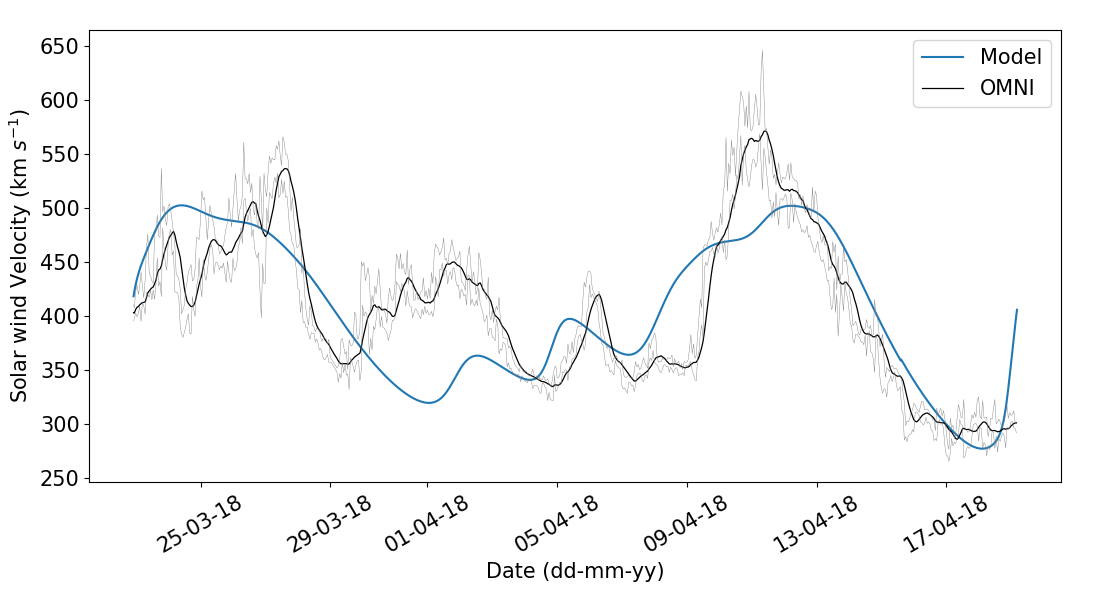}}%
        \subcaptionbox{\label{Contour_CR2202}}[0.395\textwidth]{\includegraphics[trim={0 0 0 0cm},clip,width=0.395\textwidth]{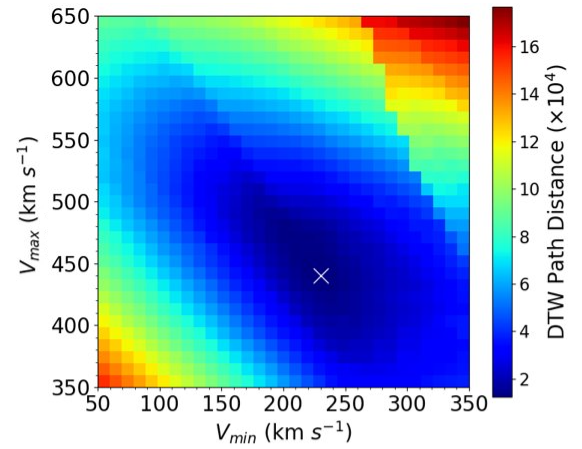}}
        \vskip\baselineskip
        \subcaptionbox{ CR2150 \label{Best_Fit_CR2150}}[0.595\textwidth]{\includegraphics[trim={0 0 0 0cm},clip,width=0.595\textwidth]{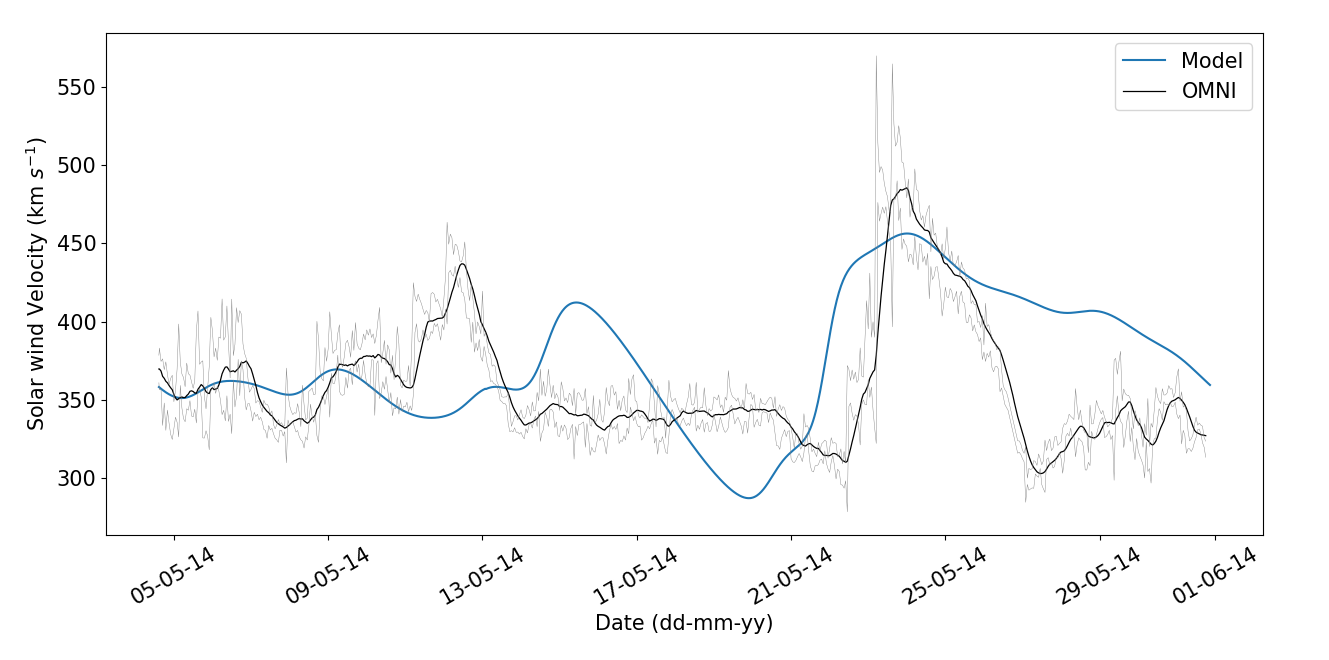}}
        \subcaptionbox{\label{Contour_CR2150}}[0.395\textwidth]{\includegraphics[trim={0 0 0 0cm},clip,width=0.38\textwidth]{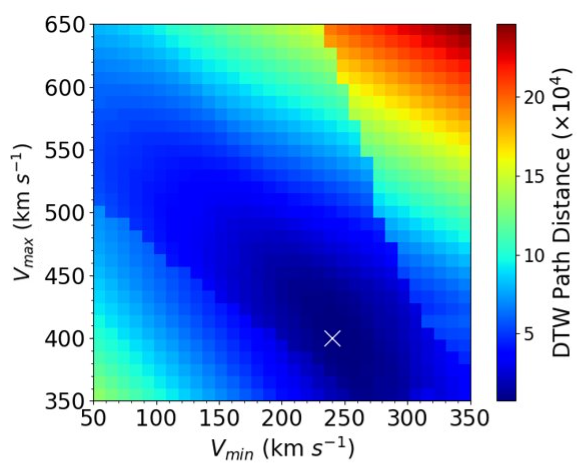}}

        \caption[ The average and standard deviation of critical parameters ]
        {(a) Comparison of model and OMNI solar wind velocities for CR 2202 for the optimal fit in terms of minimum DTW path distance (marked by white cross in figure \ref{Contour_CR2202}), with $V_{min}$ and $V_{max}$ set to 230 and 440\kms\ respectively. b) Contour plot showing the DTW path distance as a function of $V_{min}$ and $V_{max}$ for CR 2202. c) Same as (a), but for CR 2150, with $V_{min}$ and $V_{max}$ set to 240 and 400\kms\ respectively (represented by white cross in figure \ref{Contour_CR2150}). d) Same as (b), but for CR2150.} 
        \label{CR2202_CR2150}
    \end{figure*}

Here we apply the Tomography/HUXt method to two different periods during solar cycle 24. The model is applied to 2014 May (CR2150, near solar maximum) and 2018 March (CR2202, at the start of the current solar minimum). The tomography maps for these two dates are shown in figure \ref{tomoresults}. Figures \ref{Best_Fit_CR2202} and \ref{Best_Fit_CR2150} show the comparison of modelled solar wind velocity at Earth and \insitu\ data for CR2202 and CR2150 respectfully, with optimised velocity parameters which are gained from the contour plots seen in figure \ref{Contour_CR2202} and \ref{Contour_CR2150}.

For CR2202, Figure \ref{Contour_CR2202} demonstrates a minimum optimal DTW path distance at a $V_{max}$ value of 440\kms\ and a $V_{min}$ value of 230\kms\ . For this period the model and observation data agree well. However, one disparity is present between 2018 March 29 -2018 April 4, where the \insitu\ data shows a region of higher solar wind velocity which is not present in the modelled data. The mean velocity difference between \insitu\ and modelled data along the optimal DTW warped path is 0.49\kms. The time domain also shows an acceptable agreement, with a mean time difference of 1.12 days.     

CR2150 spans an active period close to the height of solar maximum. Figure \ref{Best_Fit_CR2150} shows a significant fast solar wind stream at 2014 May 22-25 which is seen in both model and measurement. However, there is a large discrepancy between a peak seen in the model at 2014 May 14-16 and one seen in measurement 3 days earlier. Figure \ref{Contour_CR2150} demonstrates a minimum optimal DTW path distance at 400 and 240\kms\ for $V_{max}$ and $V_{min}$ respectively. The velocity difference has a mean value of -0.9\kms. These values are close to the previous case studies. However the time difference is much greater in comparison to previous Carrington rotations with a mean of -1.6 days and a standard deviation of 2.3 days. Such a deteriorated agreement is likely due to the increased solar activity during this time, which both disrupts the tomography process, makes the time-independent static tomography approach less valid, and increases the chance of CMEs in the \insitu\ measurements. 

Table \ref{Table:Validation} shows a statistical comparison between both optimised tomography/HUXt and MAS/HUXt models with \insitu\ data for CR2150 and CR2202. For CR2150, the tomography/HUXt model combination yields a lower MAE (38.44 \kms) compared to the MAS/HUXt model (40.12 \kms). The MAE for CR2202 offers a significant 32\% reduction for the tomography/HUXt model (39.31 \kms), compared to that of the MAS/HUXt model (57.93 \kms). For both periods, the tomography/HUXt model shows a smaller DTW path distance and a smaller SSF than MAS/HUXt model. 

\begin{table}[h!]
\begin{tabular}{||c c c c c c c c c||} 
 \hline
 CR & IB & $V_{max}$ & $V_{min}$ & $\Delta$ T & $\Delta$ V & MAE & DTW Dist & SSF\\ [0.5ex]
 & &(km s$^{-1}$) & (km s$^{-1}$) &(Days) & (km s$^{-1}$) & (km s$^{-1}$) & $(\times10^4)$ &\\
 \hline\hline
  2150 & Tom & 400 & 240 &  -1.57 $\pm$ 2.31 & -0.80 $\pm$ 9.83 & 38.44 & 0.79 & 0.42\\
  2150 & MAS & 390 & 270 & 3.61 $\pm$ 2.19 & 0.78 $\pm$ 16.10  & 40.12 & 1.78 & 0.95\\
 \hline
  2202 & Tom & 440 & 230 & 1.12 $\pm$ 1.69 & 0.50 $\pm$ 14.92 & 39.31 & 1.25& 0.34\\
  2202 & MAS & 420 & 290 & 2.98 $\pm$ 3.21 & -3.34 $\pm$ 28.09 & 57.93 & 2.39 & 0.66\\
 \hline
\end{tabular}
\caption{Statistics of HUXt model run with both tomography and MAS inner boundaries, with the type of inner boundary indicated by the IB column. The CR column gives the Carrington rotation number, the $V_{max}$ column gives the tomography $V_{max}$ and the MAS $VM_{max}$ parameters, and the $V_{min}$ column gives the tomography $V_{min}$ and the MAS $VM_{min}$ parameters.}
\label{Table:Validation}

\end{table}

The model solar wind speeds for Carrington rotations 2150 and 2202 show a more gradual transition between high and low solar wind velocities than is seen in the observational data. This is likely due to the smoothness of the density given by the tomography, and the upwind dependence of HUXt. The significant velocity overestimation of \app75-100\kms\ seen between 2014 May 27-29 (figure \ref{Best_Fit_CR2150}) could be due to the upwind dependence of HUXt, or due to slow wind from equatorial coronal holes. Parker solar probe (PSP) has detected low density - low velocity solar wind structures at distances close to the sun that are thought to originate from equatorial coronal holes \citep[see:][]{Bale2019,Kasper2019}. This specific structure of solar wind is in direct contradiction to the simple inverse relationship of equation \ref{P_V_Empricial_Model}, which assumes a consistent low density - high velocity relationship. This will result in an overestimation of solar wind velocities at the inner boundary and therefore at Earth. A low density - low velocity structure could well be the explanation of the velocity overestimation of 2014 May 27-29 as this region maps back to approximately 150\de longitude which is a section in between two coronal holes (see figure \ref{tomoresults} a). However, this is an area with obvious key implications for solar wind forecasting that demands further research.

\subsection{A Persistence Approach} \label{persistence}

\begin{figure*}[h!]
        \centering
        \subcaptionbox{CR2203 \label{Persistence_2203}}[0.495\textwidth]{\includegraphics[trim={0cm 0 0 0cm},clip,width=0.485\textwidth]{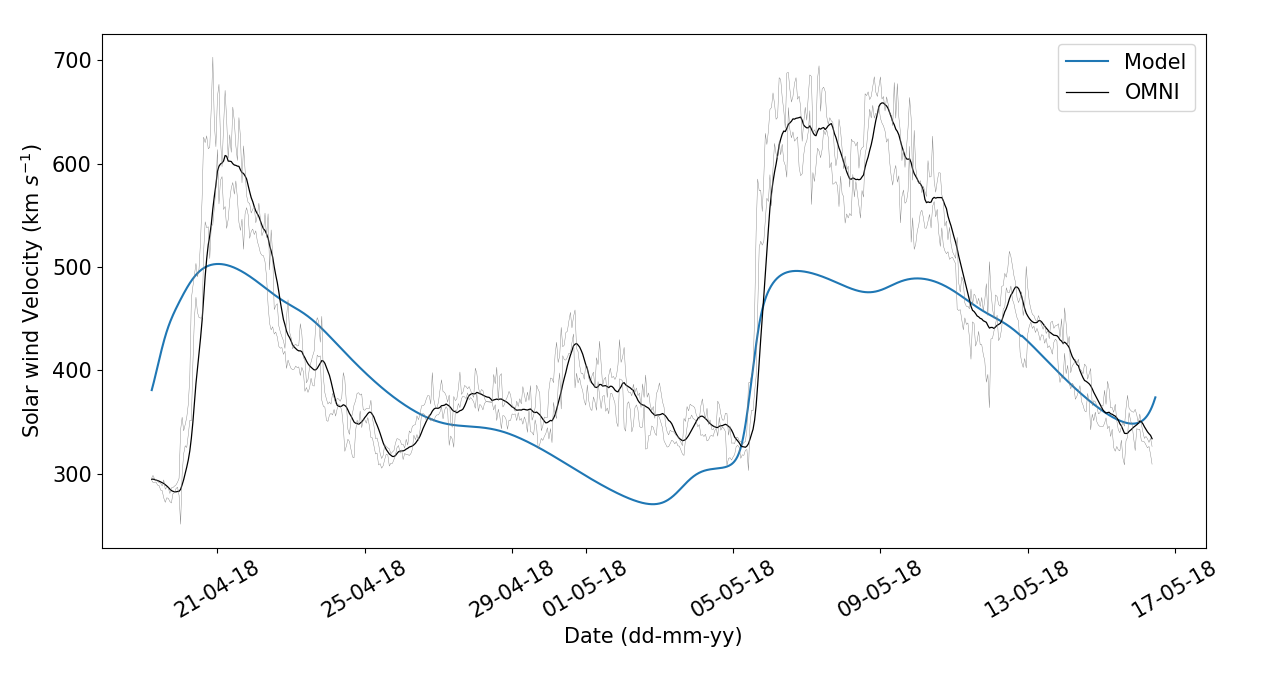}}%
        \subcaptionbox{CR2151 \label{Persistence_2151}}[0.495\textwidth]{\includegraphics[trim={0cm 0 0 0cm},clip,width=0.495\textwidth]{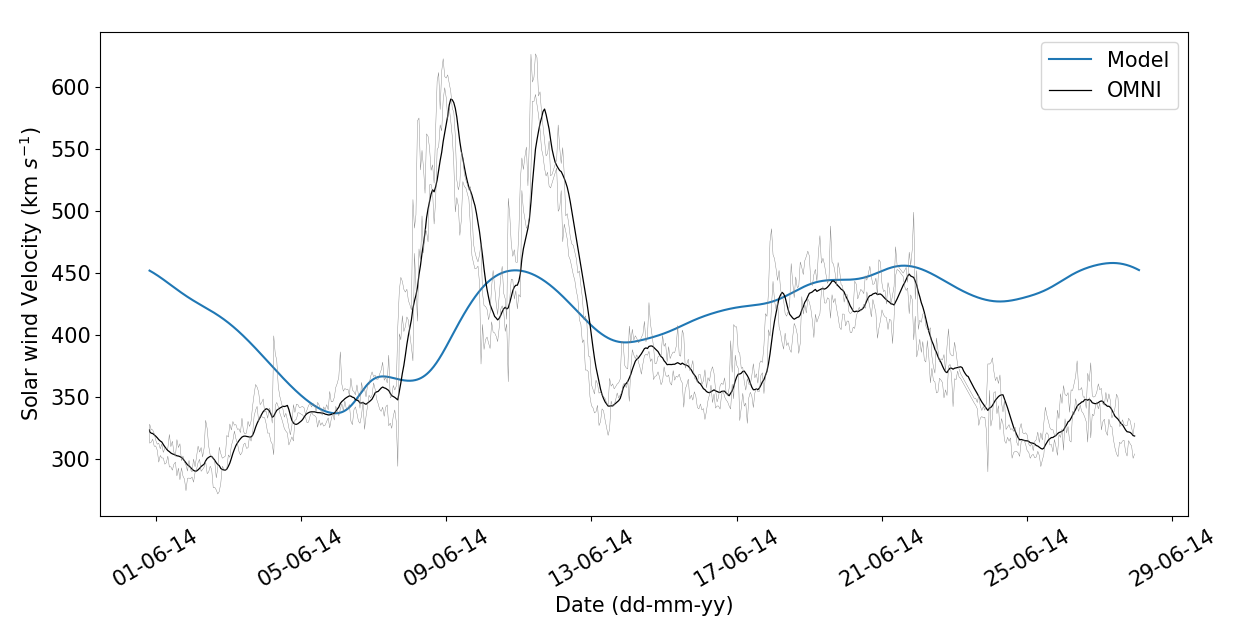}}
        
        \caption[a   ]
        {Comparison of predicted solar wind conditions of a) CR2203 ($V_{max}$ and $V_{min}$ of 440 and 230\kms\ respectively) and b)CR2151 ($V_{max}$ and $V_{min}$ of 400 and 240 \kms\ respectively). The $V_{max}$ and $V_{min}$ values are that of the optimised velocity terms of the Carrington rotation prior (obtained in section \ref{Section:application to other dates}.)    } 
        \label{Persistence}
    \end{figure*}

In this section a persistence based approach is adopted in order to attempt to predict solar wind velocities in a realistic operational context. Unlike a traditional persistence model which predicts a near-exact repetition of the observed ambient solar wind conditions for the solar rotation prior \citep[see][]{owens2013}, in this case the inner boundary condition is updated. The coronal densities (and therefore $n_{max}$ and $n_{min}$ terms in eq. \ref{P_V_Empricial_Model}) are extracted as described in section \ref{section:Den-Vel model}. However, this inner boundary condition will not undergo the exhaustive optimisation process described in section \ref{Vsolution}, but instead use the optimised $V_{max}$ and $V_{min}$ terms unchanged from Carrington rotation prior. For example, figure \ref{Persistence_2203} shows the comparison between \insitu\ data and the forecast data for CR2203 with $V_{max}$ and $V_{min}$ values of 440 and 230\kms\ which are the optimised parameters for CR2202. Likewise, figure \ref{Persistence_2151} compares \insitu\ data with forecast data for CR2151 with $V_{max}$ and $V_{min}$ of 400 and 240 \kms\ (optimised values for CR2150).

Figure \ref{Persistence_2203} shows a relatively good agreement with the \insitu\ data in terms of time of arrival of the fast solar wind streams (see table \ref{Table:Persistence_stats} for statistical analysis). However, there are disparities in the magnitude of the peaks. For example, in the first peak (seen in the \insitu\ data at 2018 April 21-22) the \insitu\ data predicts a velocity of \app 600\kms\ whereas, the model predicts a solar wind velocity of \app 500\kms. The second peak (2018 May 6-11) shows a similar difference. This suggests that the $V_{max}$ term is underestimated. Likewise, the slower more settled solar wind present in between these two peaks (2018 April 25 - 2018 May 5) is underestimated by the model, suggesting that the  $V_{min}$ term is also underestimated. Table \ref{Table:Persistence_stats} shows an SSF value of 0.65 for CR2203. This shows that a persistence approach to the tomography based model can yield more accurate results compared to a single mean value.

Figure \ref{Persistence_2151} shows a weaker agreement between the modelled and \insitu\ data. The profiles are different, with the \insitu\ data demonstrating a double peak between the dates of 2018 June 8-13, whereas the model predicts a single peak around this time. The magnitude of this peak also disagree by \app 150\kms. This again suggests that the $V_{max}$ term is underestimated. Table \ref{Table:Persistence_stats} shows generally weaker statistics for CR2151 compared to that of CR2203. A SSF value of 1.20 infers that a mean solar wind velocity across the full time period would yield a smaller DTW distance, and potentially better predictions of solar wind velocity compared to a persistence-based approach during this time period. During periods near solar maximum (such as CR2151) we would expect a persistence-based approach to yield worse results. This is due both to the coronal state changing at a significantly faster rate than during solar minimum, and the tomography reconstruction failing to accurately map the true coronal density. Therefore, the inner boundary condition will differ significantly to the physical state of the solar corona. This highlights the need for a time-dependent tomography approach in an operational context.

\begin{table}[h!]
\begin{center}
\begin{tabular}{||c c c c c c c ||} 
 \hline
 Car rot. &$V_{max}$ & $V_{min}$ & MAE  & Pearson & DTW Path Dist & SSF  \\
 & (km s$^{-1}$)& (km s$^{-1}$)& (km s$^{-1}$) & & $(\times10^4)$ & \\

 \hline\hline
  2203 & 440 & 230 & 63.74 & 0.736 & 3.77 & 0.65\\
 \hline
  2151 & 400 & 240 & 62.067 & 0.09 & 4.42 & 1.20 \\ 
 \hline
\end{tabular}
\caption{Statistical analysis of the comparison between the tomography/HUXt model with \insitu\ data using optimal velocity parameters gained for the Carrington rotation prior. }
\label{Table:Persistence_stats}
\end{center}
\end{table}

Overall, here we show that a persistence model could potentially be used in an operational context as a worst case scenario, without updating the $V_{min}$ and $V_{max}$ parameters. However, the model certainly loses accuracy when deployed in this fashion especially during periods near solar maximum. This stresses the need of either a time-dependent boundary condition, or a more complex, global relationship between coronal electron density and solar wind velocity at a height of 8\Rs. Both of these issues are the focus of our current efforts.

\section{An Ensemble Approach} \label{Ensemble}
The high efficiency of the HUXt model allows an ensemble approach to estimate the uncertainty in model outputs based on selected uncertainties at the inner boundary. The tomography density distribution, as shown in figures \ref{Tomo_map_CR2210} and \ref{tomoresults} show thin elongated structures that tend to lie longitudinally, and can be very narrow in latitude. Therefore, a small error in the distribution given by the tomography method, or small latitudinal deviance of the solar wind during propagation to Earth, can significantly alter the inner boundary condition and the predicted solar wind conditions at Earth. An ensemble approach is a straightforward way of investigating and quantifying the effect small variations in latitude of the extracted tomographical data at the inner boundary can have on the resulting solar wind velocities at Earth. Another uncertainty to which we can apply the ensemble approach is the choice of $V_{max}$ and $V_{min}$. Here, we use the map of DTW cost function arising from the exhaustive search of $V_{max}$ and $V_{min}$ values to define a range of velocity terms at the inner boundary for 2014 May (CR2150) and 2018 March (CR2202).

\subsection{Latitudinal dependence}

Figure \ref{tomoresults} shows two density maps that demonstrate two extremes of solar activity. CR2150, as shown in figure \ref{tomoresults}a, demonstrates a complicated density distribution with many high density streamers positioned at a wide range of latitudes, and which span across the equator. Figure \ref{tomoresults}b shows a quiet solar corona where the streamer belt is longitudinally aligned near the equator, and a more uniform low electron density at higher latitudes. The main density enhancements are found exclusively in the equatorial region.

\begin{figure}[h!]
        \centering
        
        \subcaptionbox{CR2150\label{CR2150_lat}}[0.8\textwidth]{\includegraphics[trim={0 0 0 0cm},clip,width=0.8\textwidth]{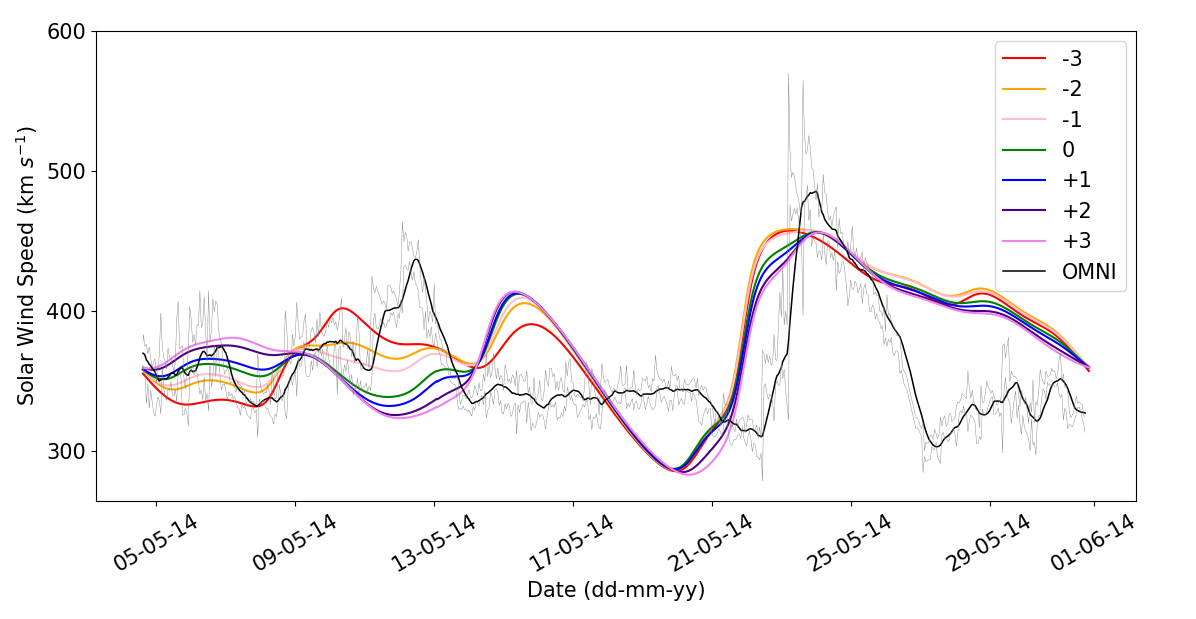}}%
        \vskip\baselineskip
        \subcaptionbox{CR2202\label{CR2202_lat}}[0.8\textwidth]{\includegraphics[trim={0 0 0 0cm},clip,width=0.8\textwidth]{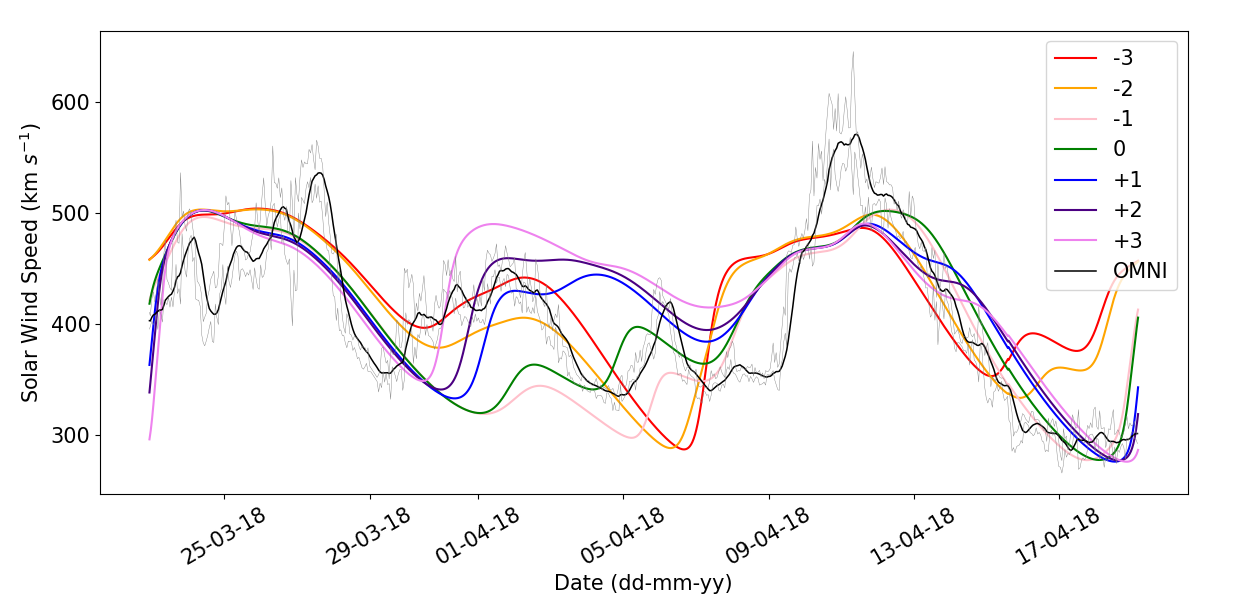}}
        
        \caption[ The average and standard deviation of critical parameters ]
        { Comparison of \insitu\ data with the results of multiple tomography/HUXt model runs for (a) CR2150 and (b) CR2202, with the latitude of the inner boundary condition varied in one degree increments between $\pm$3\de\ from the latitude of Earth.} 
        \label{Latitude_comparison}
    \end{figure}

A model run was conducted as described in section 3.3 with $V_{max}$ and $V_{min}$ remaining fixed at the optimised values of 400 and 240\kms\ for CR2150 and 440 and 230\kms\ for CR2202. We adjust the inner boundary velocity profile by extracting densities from the tomography map with $\pm$3\de\ of the latitude of Earth with 1\de\ increments. This range of latitudes were chosen with the aim of comfortably covering the latitudinal movement of Earth during one full Carrington rotation, and, more importantly, the unknown drift of the solar wind in latitude between the Sun and Earth. Comparisons between the resulting model and measured solar wind conditions at Earth are shown in figure \ref{Latitude_comparison}. The largest variations in the ensemble velocities during CR2150 (see Figure \ref{CR2150_lat}) are present before 2014 May 13. During this time we find that the relatively small latitude variation of $\pm$3\de\ leads to a wide variation in velocity. For example, the largest velocity range is \app\ 50\kms\ observed on 2014 May 10. For the remainder of the period, varying the latitude leads to only a small variation in velocity (\app\ 20\kms). This can be explained by examining the density map of figure \ref{tomoresults}a. The solar wind streams at Earth, corresponding to the early part of the period, map back approximately to the region bounded by the red box labelled A in the tomography map. This location was calculated by considering the solar wind speed, the distance from Earth to the tomography map location, and the solar rotation rate. This region spans the boundary between a high density streamer to the north, and a low density coronal hole to the south. Therefore, the northern  latitudes in figure \ref{CR2150_lat}, yield considerably slower solar wind velocities compared to that of southern latitudes during this time. Thus varying the latitude by small values can lead to large changes in the density profile, and therefore the inner boundary velocity.

All ensemble model runs fail to match the steep decrease in  velocity during 2014 May 25. The coordinates of this decrease near Earth map back to the region bounded by the red box labelled B in the tomography map. This is a narrow region of low density lying between longitudinally extended regions of higher density to the north and south. Either the tomography map is incorrect in this small area, and that this region should contain a higher density and thus map to slow velocities, or the simplistic inverse linear relationship  mapping densities to velocities, as given by equation \ref{P_V_Empricial_Model} fails in this region. If the latter, then the relationship should not be linear, and should give highest velocities only for the very lowest density features in the tomography map. Current efforts are focused on the investigation of an improved relationship between coronal electron density and solar wind velocity at 8\Rs.

All of the ensemble members consistently predict a higher velocity peak at a date of 2014 May 14-16. This feature is not on seen on these dates in the \insitu\ data. This peak maps back to the small low density feature at Carrington longitude 270\de\ as seen in the equatorial region of the tomography map seen in figure \ref{tomoresults}a. We note that the density of this feature is low, but not at the minimum densities estimated for coronal holes seen in this map and others. Conversely, the peak at intermediate velocities seen in the data around 2014 May 12 is not present on this date in any of the model runs. This feature maps back to a longitude of around 305\de, close to the boundary between the low-density region at 270\de.  It is likely that the velocity peak seen in all ensemble members (2014 May 14-16) is the same peak seen in the \insitu\ data during 2014 May 12. The amplitude of both peaks is comparable. There are several reasons why the model predicted a later time of arrival at Earth for this specific structure, including the modelling limitations of HUXt (e.g. the acceleration parameter), or a defect in the tomography map. 

\begin{figure}[h!]
    \centering
    \includegraphics[width=0.7\textwidth]{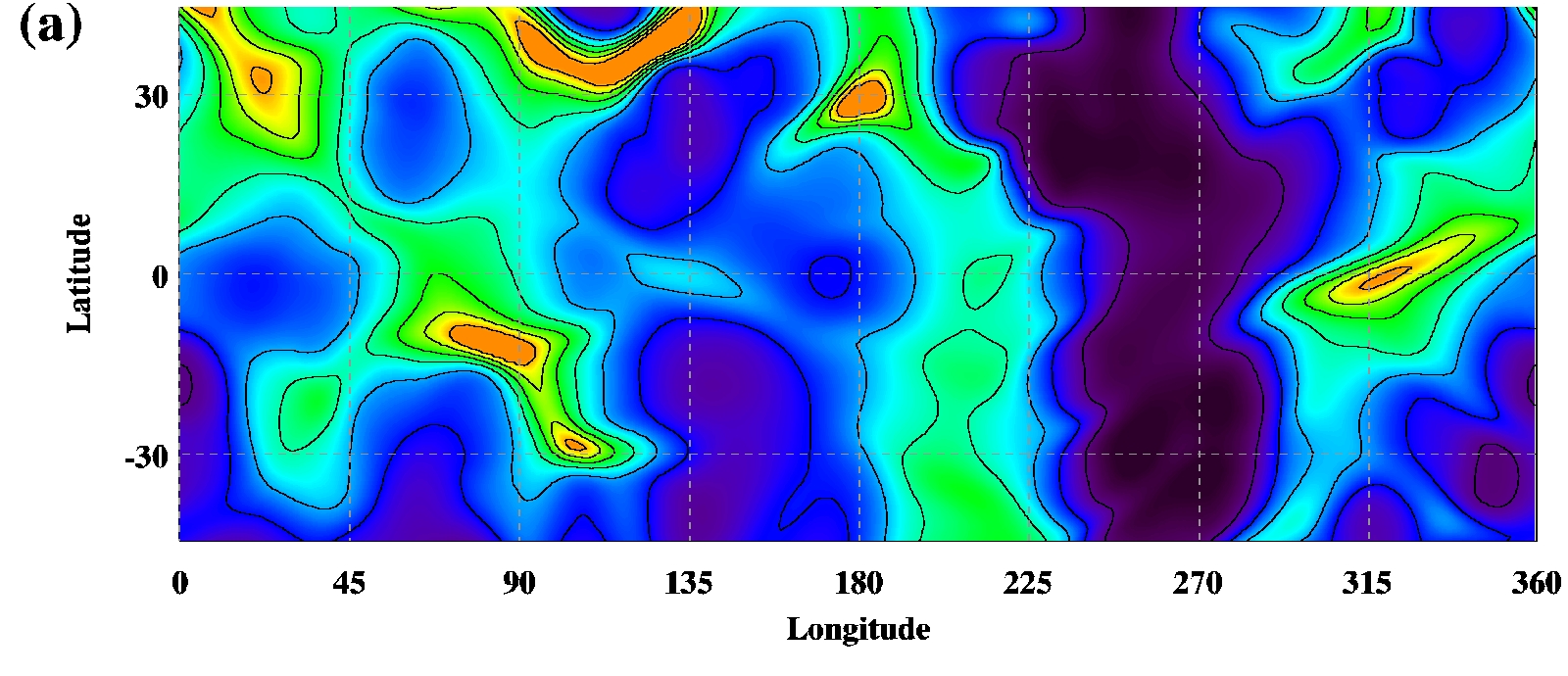}
    \includegraphics[width=0.7\textwidth]{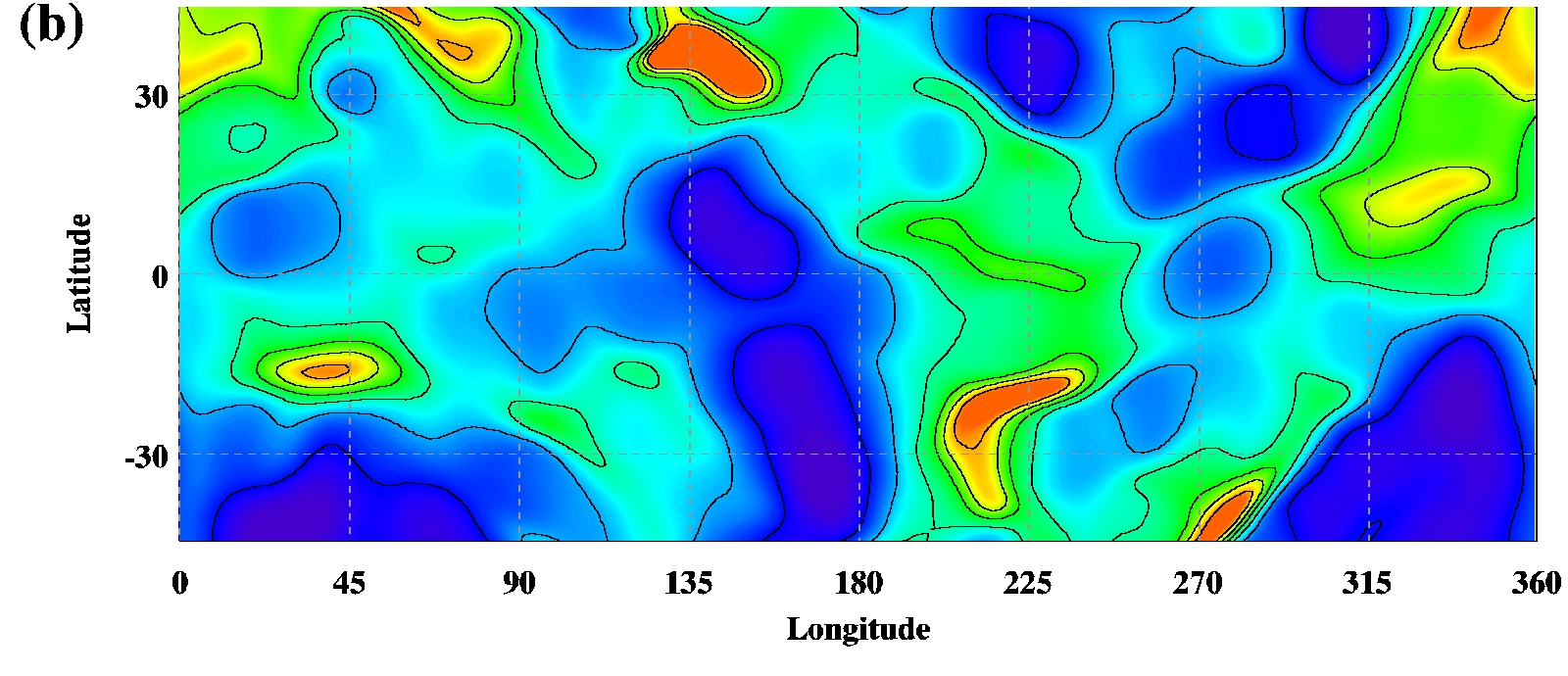}
    \includegraphics[width=0.7\textwidth]{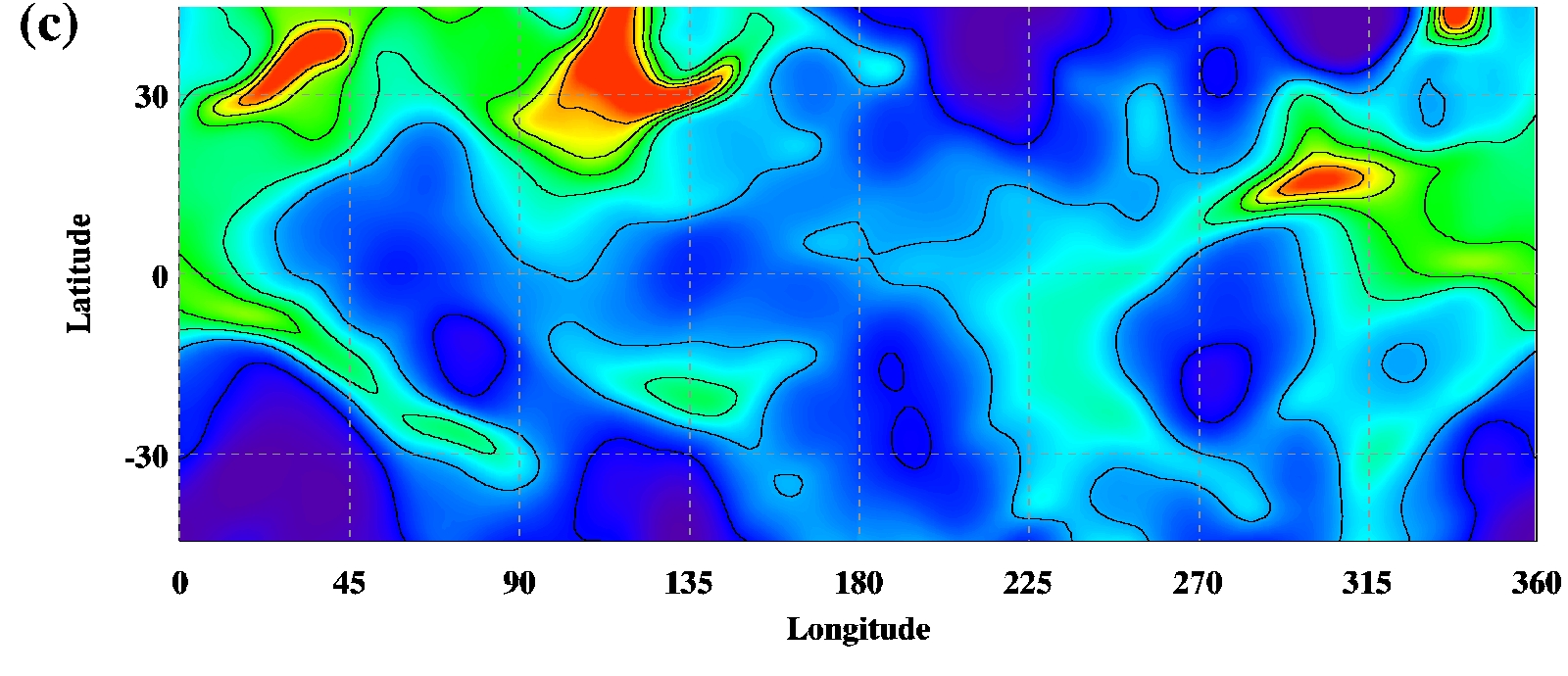}
    \caption{The rapidly changing density of the corona between (a) CR2149, (b) CR2150, and (c) CR2151.}
    \label{tomoresults2}
\end{figure}

Figure \ref{tomoresults2} shows how the density of the corona changes in the time of just three Carrington rotations. During CR2149, shown in figure \ref{tomoresults2}a, there is a large and very low density coronal hole spanning the equator and reaching to high latitudes. There is a clearly defined western boundary to this coronal hole at longitude 280\de\ near the equator. By CR2150, as shown in figure \ref{tomoresults2}b, the coronal hole has greatly reduced in size and does not reach the low densities of the previous rotation. Consequently, the western boundary is not clearly defined. The following rotation in figure \ref{tomoresults2}c shows the western streamer encroaching into the region previously occupied by the coronal hole, and the coronal hole limited to southern regions. This rapid change in structure is the best explanation for the differences between the model and \insitu\ measurements between dates 2015 May 12 and 18, and shows that a time-dependent inner boundary becomes critical during solar maximum.

The results for CR2202 in figure \ref{CR2202_lat} show that the variability in solar wind prediction is largest between 2018 April 1 to 8, where the model velocities vary by more than 100\kms. This region maps back to the white boxed region in figure \ref{tomoresults}b. This region spans the northern boundary of a high-density streamer, thus small variations in latitude lead to large variations in density and velocity. This result shows that even during quiet periods, large uncertainties can arise from small deviations in latitude. This is a significant problem for solar wind forecasting, particularly considering that the high-density streamer belt may actually be narrower than that reconstructed using tomography. The measured fast wind peak during 2018 April 11 reaches speeds of almost 600\kms. The models over all latitudes consistently underestimate this peak. This feature maps back to the low density equatorial region near Carrington longitude 145\de. This is likely due to the tomography map overestimating the density at this point, or a flaw in the oversimplistic relationship between density and velocity, or a combination of both.

\subsection{Velocity dependence}

\begin{figure*}[h]
        \centering
        
        \subcaptionbox{CR2150 \label{Heat_Map_CR2150}}[0.495\textwidth]{\includegraphics[width=0.495\textwidth]{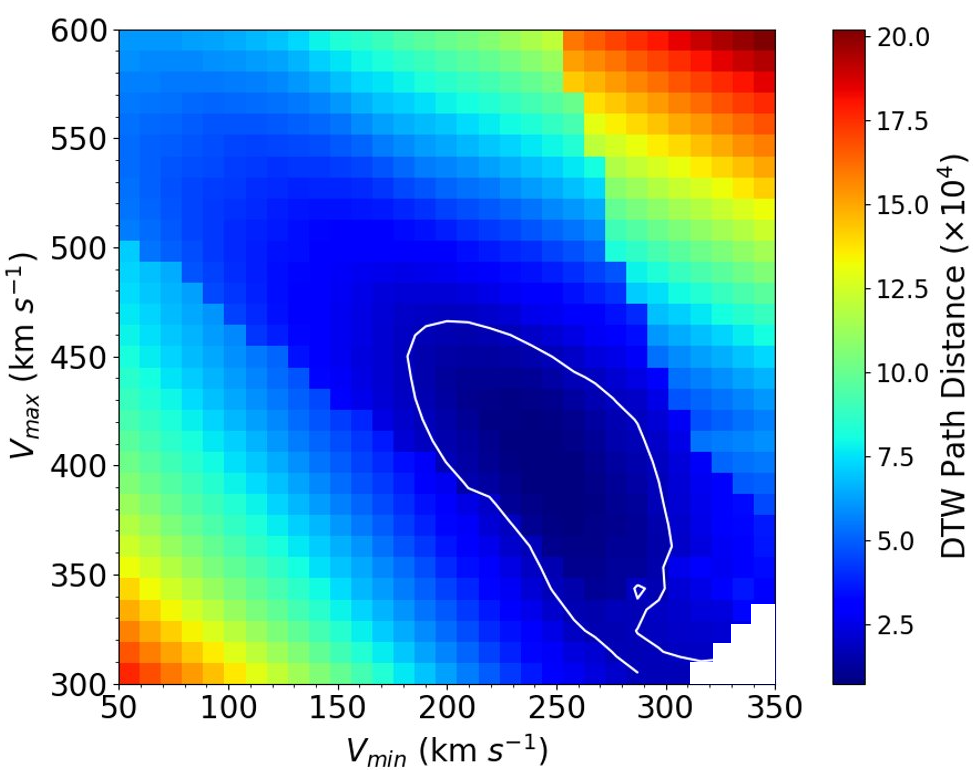}}%
        \subcaptionbox{CR2202 \label{Heat_Map_CR2202}}[0.495\textwidth]{\includegraphics[width=0.495\textwidth]{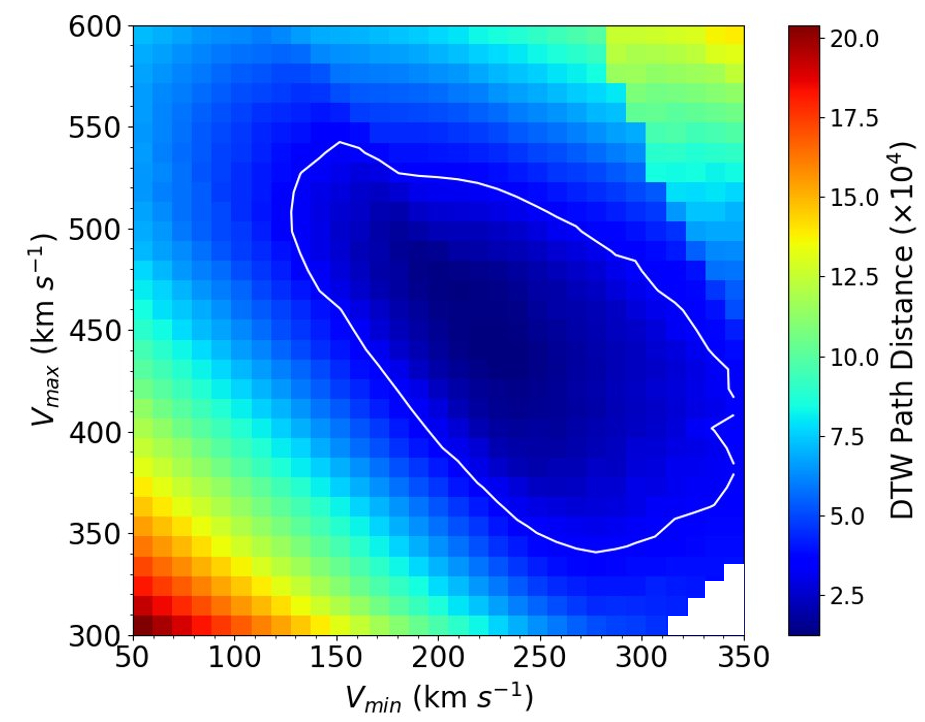}}
        
        \caption[ The average and standard deviation of critical parameters ]
        { Contour plots of the DTW path distance as a function of minimum and maximum inner boundary velocity for Carrington rotations (a) 2150, and (b) 2202. The white contour bounds the area that yields a DTW path distance of under the threshold chosen (SSF $\le$ 0.95) and the white pixels in the lower right corner represent invalid model parameters where $V_{min}$ \textgreater $V_{max}$. Note that the $V_{max}$ range is altered to 300-600\kms\ in this plot.}
        \label{Heat_maps}
    \end{figure*}

Figure \ref{Heat_maps} shows the DTW path distance as a function of $V_{max}$ and $V_{min}$, used to select the optimal parameters. In order to create a velocity-based ensemble, we wish to identify a region within this parameter space where the DTW distance is lower than a set threshold.      

Figure \ref{Heat_maps} shows that there is a selection of velocity values that give an acceptable fit for \insitu\ data, defined as SSF $\le$ 0.95. This range of velocity magnitudes defines an uncertainty which is incorporated into the ensemble. The velocity terms for each ensemble member are randomly generated from a normal distribution, with a mean set at the optimal velocity parameters, and a range within three standard deviations of the values that yield a SSF of 0.95 or lower. These regions are highlighted by the white contour in figure \ref{Heat_maps}. For example, CR2150 has mean (optimal) values of 400 and 240\kms\ and a standard deviation of 20\kms\ for $V_{max}$ and $V_{min}$ respectively. For CR2202, mean velocity values are 440 and 230\kms\ with a standard deviation of 30\kms. Care was taken in order to ensure $V_{max}$ term was greater than the $V_{min}$ term for every ensemble member.

\begin{figure}[h!]
        \centering
        \subcaptionbox{CR2150\label{CR2150_V_changes}}[0.8\textwidth]{\includegraphics[trim={0 0 0 0cm},clip,width=0.8\textwidth]{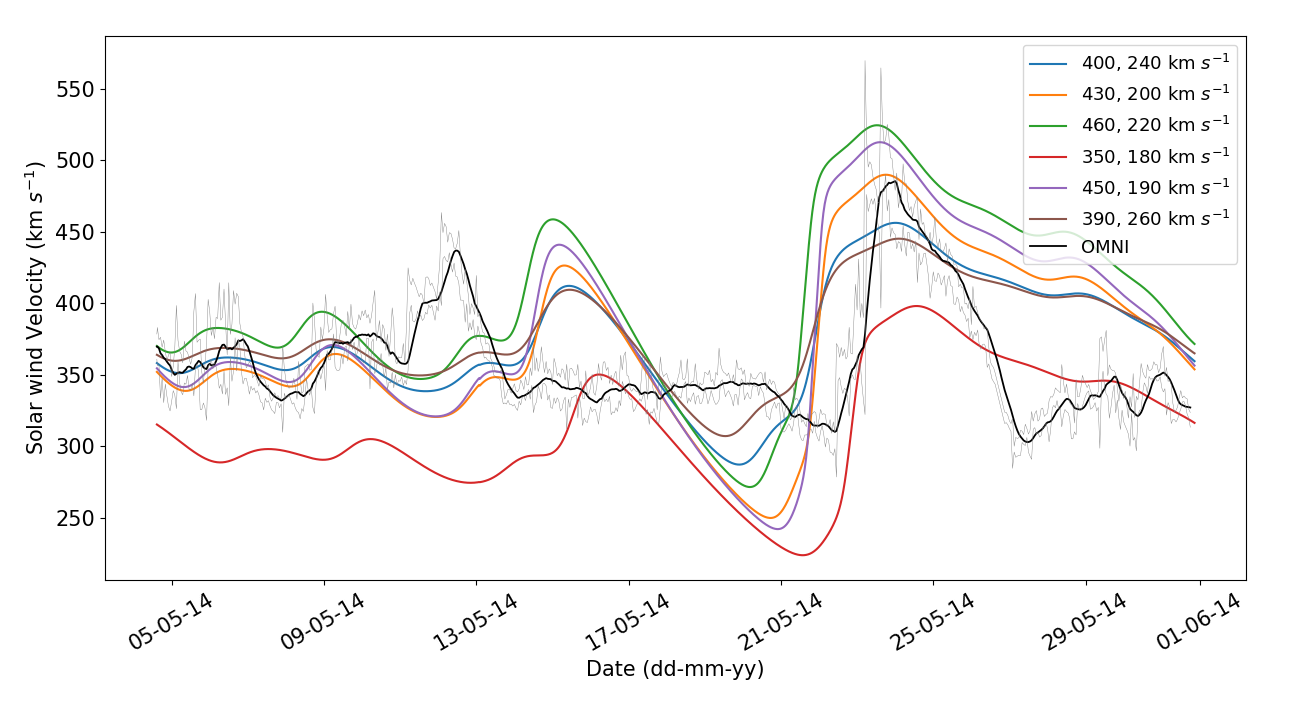}}%
        \vskip\baselineskip
        \subcaptionbox{CR2202\label{CR2202_V_changes}}[0.8\textwidth]{\includegraphics[trim={0 0 0 0cm},clip,width=0.8\textwidth]{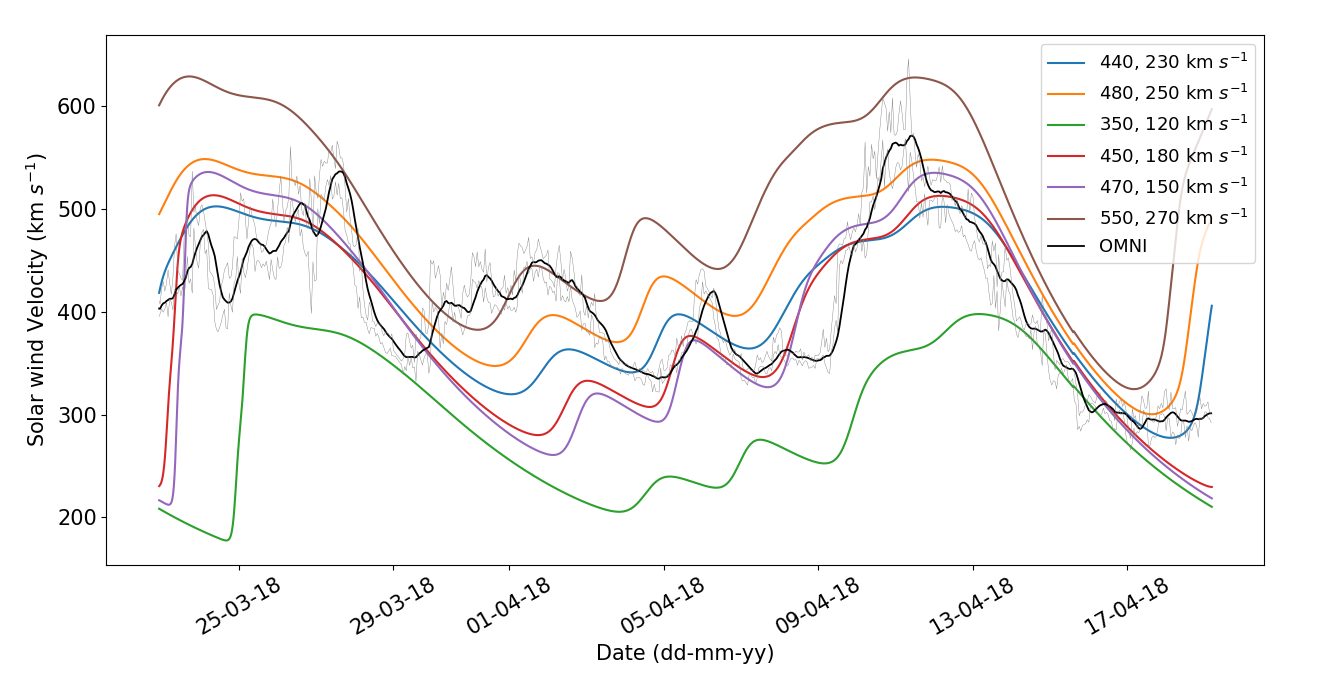}}
        
        \caption[ The average and standard deviation of critical parameters ]
        { Comparison of \insitu\ data with the results of multiple tomography/HUXt model runs for (a) CR2150 and (b) CR2202, with a range of $V_{max}$ and $V_{min}$ combinations.} 
        \label{Velocity_comparison}
    \end{figure}

Figure \ref{Velocity_comparison} compares \insitu\ data with seven tomography/HUXt outputs each with a different combination of $V_{max}$ and $V_{min}$ magnitudes for CR2150 and CR2202. This highlights the sensitivity of the output to the velocity range. An underestimated $V_{max}$ value will affect the magnitude of the solar wind peaks predicted at Earth, but will also cause the fast solar wind peak to arrive later. For example in figure \ref{Velocity_comparison}a, the model with $V_{max}=350$\kms\ (red) shows a 2-3 day delay in time of arrival of fast solar wind at 2018 May 26, compared with a model run with $V_{max}=460$\kms\ (green). Both these terms dictate how rapid the transition between fast and slow solar wind occur.

\subsection{Ensemble Results}
\begin{figure}[h!]
        \centering
        
        \subcaptionbox{CR2150\label{CR2150_ensemble}}[0.8\textwidth]{\includegraphics[width=0.8\textwidth]{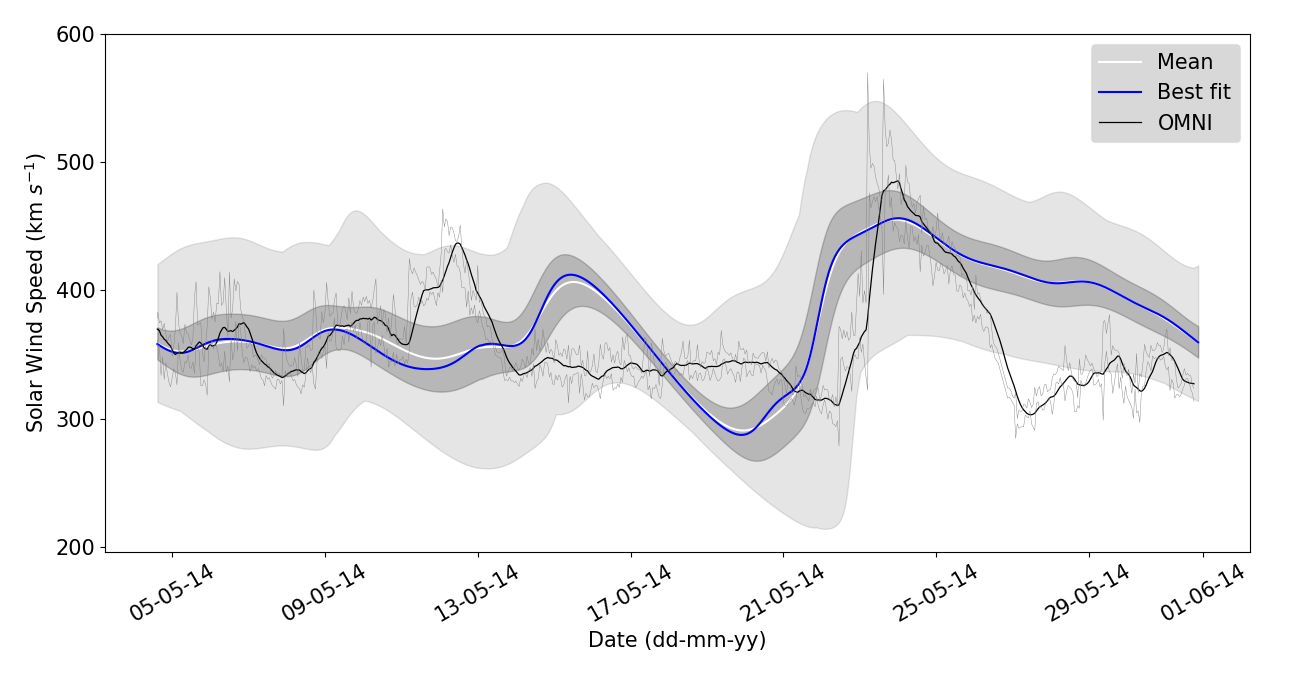}}%
        \vskip\baselineskip
        \subcaptionbox{CR2202\label{CR2202_ensemble}}[0.8\textwidth]{\includegraphics[width=0.8\textwidth]{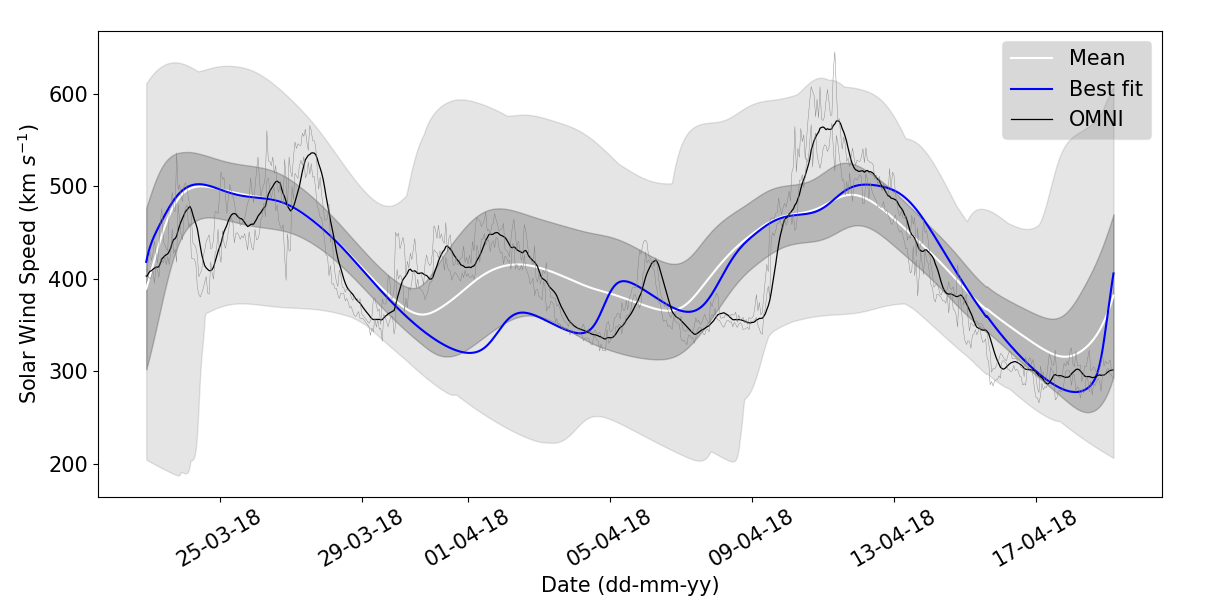}}

        \caption[ The average and standard deviation of critical parameters ]
        {Results of the ensemble for a) CR2150 and b) CR2202 with the mean predicted solar wind velocity at each time step shown as white, the light grey area representing all predicted solar wind velocities, and the darker grey representing one standard deviation from the mean. The non-ensemble model results with the optimal velocity values at the Carrington latitude of Earth is the blue line. }
        \label{Ensemble_results}
    \end{figure}

We create an ensemble with 10,000 unique inner boundary conditions generated with randomly selected pairs of $V_{max}$ and $V_{min}$ taken from a normal distribution around the optimal velocities, and a line of latitude randomly selected between +/- 3 degrees from Earth's latitude at the initiation of the Carrington rotation. Once generated, the 10,000 inner boundary conditions were used for the HUXt model and the results of solar wind velocities at Earth recorded. The results of the ensemble are shown in figure \ref{Ensemble_results}. A choice of 10,000 ensemble runs was a compromise between quantifying the sensitivity of the results to the parameter uncertainties with a sensible computation time.
    
Figure \ref{CR2150_ensemble} shows a strong correlation between the mean of the ensemble runs (white line) and the optimal HUXt model for the true latitude of Earth (blue line), with a maximum velocity difference of 12\kms. The standard deviation of the ensemble around the mean is shown as the dark grey region, and the average standard deviation across the full time series is 20.06\kms. The most significant deviations between the \insitu\ data and the model data lie at three intervals centered on 2014 May 12, 2014 May 15 (as discussed in section \ref{Section:application to other dates}) and 2014 May 27. The final disparity shows a steep transition between the fast and slow solar wind (seen around 2014 May 28 in the \insitu\ data) that is not predicted by any of the 10,000 ensemble model runs.

CR2202, as seen in figure \ref{CR2202_ensemble}, also demonstrates a good agreement with the \insitu\ data with the maximum velocity difference between the mean and the \insitu\ data being 79\kms. CR2202 shows a greater variability of the ensemble model runs (as shown by the height of both dark grey and light grey regions) compared that of CR2150, with the average standard deviation across the full time-series being 41.7\kms. This is approximately double that of CR2150. It would be expected that value is larger than that for CR2150 due to the larger standard deviation entered into the velocity terms (20 \kms\ and 30\kms\ for CR2150 and CR2202 respectively).

\section{Discussion and conclusions}
\label{conclusions}
A new inner boundary condition for solar wind heliospheric models is derived from coronal electron density tomography maps. The tomography/HUXt model results give general good agreement with \insitu\ data provided by the OMNI satellite network, with a significant increase in the accuracy of solar wind velocity prediction compared to a HUXt model run with a traditional MAS derived inner boundary condition for the time periods used in this study. The time periods investigated in this study were intentionally chosen to demonstrate the model at different stages of the solar activity cycle. Given further development, a future study will explore a larger dataset. 

We can identify several aspects of the inner boundary condition and modelling that lead to inaccuracies, and are possible to address with some further work. They are:
\begin{itemize}
    \item The use of a static tomography reconstruction to represent a dynamic corona. We have recently developed a framework for providing time-dependent tomographical densities, that will help address this, although further development is needed \citep{morgan2021}. 
    \item The overly smooth reconstruction given by tomography compared to the true density. The coronal streamer belt likely consists of very narrow high-density structures \citep{morgan2010, morgan2007twistingsheets, morgan2007empiricalmodel} that are highly variable on small temporal \citep[e.g.,][]{alzate2021} and spatial scales \citep{ thernisien2006, poirier2020}. Again, developments in tomographical methods and observations can lead to reconstructions that bring us closer to these scales.  
    \item The use of a single, fixed acceleration profile for both fast and slow wind. We are developing our own upwind model that includes both velocities and densities, and includes the acceleration profile as a search parameter. This method uses iteration to improve the fit to \insitu\ measurements. Early results are promising, and may provide a constraint on acceleration. Another approach is to use a set of tomographical maps over a range of distances (4 to 10\Rs) to constrain the early acceleration profile of the slow wind, similar to that shown by \citet{morgan2020}. 
    \item The simplistic inverse relationship of equation \ref{P_V_Empricial_Model}. We are investigating the replacement of the simple inverse relationship of equation \ref{P_V_Empricial_Model} with a Sigmoid or exponential function, which gives a simple relationship, based on a small number of parameters, to model the transition from slow to fast wind as a function of the density. 
    \item Use of a standard, possibly incorrect, coronal rotation rate. Long time series of tomographical maps can give improved estimates of the variable coronal rotation rate \citep{edwards2021, Morgan2011, morgan2011b}. Near the equator, the coronal rotation rate may vary by a degree per day or more from the Carrington rate, leading to systematic longitudinal errors in solar wind model results. 
    \item The omission of coronal mass ejections (CME) that may be present in the \insitu\ data. This can be addressed to an extent using the current approach for operational forecasting: simple parameters describing a cone model of a CME can be input to HUXt, and the CME carried with the solar wind, giving an estimate of time of arrival at Earth. 
\end{itemize}

A persistence-based approach was made in order to test the operational feasibility of this model. The persistence results showed a large difference in accuracy between periods near solar minimum and maximum. The rate of evolution of the physical corona during solar maximum is such that a persistence approach will break down over timescales of a Carrington rotation. However, during solar minimum, results were more acceptable, yet less accurate than a non-persistence approach. This highlights the need for a time-dependent inner boundary condition, and a more advanced relationship between density and velocity than that given by equation \ref{P_V_Empricial_Model}.

The efficiency of the HUXt model allows an ensemble framework, which can quantify the uncertainties of the predicted solar wind velocities. The ensembles was based on sampling an appropriate range of latitudes and velocities at the model inner boundary. The ensemble results confirm that both these uncertainties have a significant effect on the model output velocities at Earth. Latitudinally narrow, longitudinally-aligned streamer belt structures can lead to high uncertainties in the output based on small latitudinal uncertainties at the inner boundary. The choice of velocity range at the inner boundary also has a large impact on the model results.

In the context of operational space weather forecasting, this paper shows that the use of a tomography-based inner boundary condition to drive the HUXt model as part of the CORTOM module of the SWEEP project for the UK Met Office is an approach that can offer certain improvements on current systems. Use of multiple models over extended periods will enable an extensive analysis of their relative performance, and the improvements described in this paper will be implemented within SWEEP over the coming years.

 \begin{acknowledgements}
We acknowledge (1) STFC grants ST/S000518/1 and ST/V00235X/1, (2) Leverhulme grant RPG-2019-361, (3) STFC PhD studentship ST/S505225/1, and (4) the excellent facilities and support of SuperComputing Wales, (5) NASA/GSFC's Space Physics Data Facility's OMNIWeb service and OMNI data, (6) Predictive Science Inc. for the MHDWeb service and MAS model outputs, and (7) University of Reading (UK) for the HUXt model. HUXt is available for download at \url{https://github.com/University-of-ReadingSpace-Science/HUXt.}. The STEREO/SECCHI project is an international consortium of the Naval Research Laboratory (USA), Lockheed Martin Solar and Astrophysics Lab (USA), NASA Goddard Space Flight Center (USA), Rutherford Appleton Laboratory (UK), University of Birmingham (UK), Max-Planck-Institut fu\"r Sonnen-systemforschung (Germany), Centre Spatial de Liege (Belgium), Institut Optique Th\'eorique et Appliq\'uee (France), and Institut d'Astrophysique Spatiale (France).
 \end{acknowledgements}

\bibliography{references2.bib}  

\begin{thebibliography}{70}
\providecommand{\natexlab}[1]{#1}
\providecommand{\url}[1]{\texttt{#1}}
\providecommand{\urlprefix}{URL }
\providecommand{\eprint}[2][]{\url{#2}}

\bibitem[{Allen et~al.(2020)Allen, Lario, Odstrcil, Ho, Jian
  et~al.}]{Allen2020}
Allen, R.~C., D.~Lario, D.~Odstrcil, G.~C. Ho, L.~K. Jian, et~al., 2020.
\newblock { Solar Wind Streams and Stream Interaction Regions Observed by the
  Parker Solar Probe with Corresponding Observations at 1 au }.
\newblock \emph{The Astrophysical Journal Supplement Series}.
\newblock 10.3847/1538-4365/ab578f.

\bibitem[{{Alzate} et~al.(2021){Alzate}, {Morgan}, {Viall}, and
  {Vourlidas}}]{alzate2021}
{Alzate}, N., H.~{Morgan}, N.~{Viall}, and A.~{Vourlidas}, 2021.
\newblock {Connecting the Low to the High Corona: A Method to Isolate
  Transients in STEREO/COR1 Images}.
\newblock \emph{\apj}, \textbf{919}(2), 98.
\newblock 10.3847/1538-4357/ac10ca, \eprint{2107.02644}.

\bibitem[{Arge and Pizzo(2000)}]{arge2000}
Arge, C.~N., and V.~J. Pizzo, 2000.
\newblock Improvement in the prediction of solar wind conditions using
  near-real time solar magnetic field updates.
\newblock \emph{Journal of Geophysical Research: Space Physics},
  \textbf{105}(A5), 10,465--10,479.
\newblock Https://doi.org/10.1029/1999JA000262.

\bibitem[{{Aschwanden}(2011)}]{aschwanden2011}
{Aschwanden}, M.~J., 2011.
\newblock {Solar Stereoscopy and Tomography}.
\newblock \emph{Living Reviews in Solar Physics}, \textbf{8}(1), 5.
\newblock 10.12942/lrsp-2011-5.

\bibitem[{Baker et~al.(2004)Baker, Daly, Daglis, Kappenman, and
  Panasyuk}]{Baker2004}
Baker, D., E.~Daly, I.~Daglis, J.~G. Kappenman, and M.~Panasyuk, 2004.
\newblock Effects of Space Weather on Technology Infrastructure.
\newblock \emph{Space Weather}, \textbf{2}(2).
\newblock 10.1029/2003SW000044.

\bibitem[{Bale et~al.(2019)Bale, Badman, Bonnell, Bowen, Burgess
  et~al.}]{Bale2019}
Bale, S.~D., S.~T. Badman, J.~W. Bonnell, T.~A. Bowen, D.~Burgess, et~al.,
  2019.
\newblock {Highly structured slow solar wind emerging from an equatorial
  coronal hole}.
\newblock \emph{Nature}.
\newblock 10.1038/s41586-019-1818-7.

\bibitem[{Berndt and Clifford(1994)}]{Berndt1994}
Berndt, D.~J., and J.~Clifford, 1994.
\newblock Using Dynamic Time Warping to Find Patterns in Time Series.
\newblock In Proceedings of the 3rd International Conference on Knowledge
  Discovery and Data Mining, AAAIWS'94, 359–370. AAAI Press.

\bibitem[{Brueckner et~al.(1995)Brueckner, Howard, Koomen, Korendyke, Michels
  et~al.}]{Brueckner1995}
Brueckner, G.~E., R.~A. Howard, M.~J. Koomen, C.~M. Korendyke, D.~J. Michels,
  et~al., 1995.
\newblock {The Large Angle Spectroscopic Coronagraph (LASCO)}.
\newblock \emph{Solar Physics}, \textbf{162}(1), 357--402.
\newblock 10.1007/BF00733434.

\bibitem[{{Butala} et~al.(2005){Butala}, {Frazin}, and
  {Kamalabadi}}]{butala2005}
{Butala}, M.~D., R.~A. {Frazin}, and F.~{Kamalabadi}, 2005.
\newblock {Three-dimensional estimates of the coronal electron density at times
  of extreme solar activity}.
\newblock \emph{Journal of Geophysical Research (Space Physics)},
  \textbf{110}(A9), A09S09.
\newblock 10.1029/2004JA010938.

\bibitem[{der Holst et~al.(2014)der Holst, Sokolov, Meng, Jin,
  W.~B.~Manchester, T{\'{o}}th, and Gombosi}]{van_der_Holst_2014}
der Holst, B.~V., I.~V. Sokolov, X.~Meng, M.~Jin, I.~W.~B.~Manchester,
  G.~T{\'{o}}th, and T.~I. Gombosi, 2014.
\newblock {ALFV}{\'{E}}N {WAVE} {SOLAR} {MODEL} ({AWSoM}): {CORONAL} {HEATING}.
\newblock \emph{The Astrophysical Journal}, \textbf{782}(2), 81.
\newblock 10.1088/0004-637x/782/2/81.

\bibitem[{Diego et~al.(2010)Diego, Storini, and Laurenza}]{Diego2010}
Diego, P., M.~Storini, and M.~Laurenza, 2010.
\newblock Persistence in recurrent geomagnetic activity and its connection with
  Space Climate.
\newblock \emph{Journal of Geophysical Research: Space Physics},
  \textbf{115}(A6).
\newblock 10.1029/2009JA014716.

\bibitem[{Doherty et~al.(2004)Doherty, Coster, and Murtagh}]{Doherty2004}
Doherty, P., A.~J. Coster, and W.~Murtagh, 2004.
\newblock {Space weather effects of October–November 2003}.
\newblock \emph{GPS Solutions}.
\newblock 10.1007/s10291-004-0109-3.

\bibitem[{{Eastwood} et~al.(2018){Eastwood}, {Hapgood}, {Biffis}, {Benedetti},
  {Bisi}, {Green}, {Bentley}, and {Burnett}}]{eastwood2018}
{Eastwood}, J.~P., M.~A. {Hapgood}, E.~{Biffis}, D.~{Benedetti}, M.~M. {Bisi},
  L.~{Green}, R.~D. {Bentley}, and C.~{Burnett}, 2018.
\newblock {Quantifying the Economic Value of Space Weather Forecasting for
  Power Grids: An Exploratory Study}.
\newblock \emph{Space Weather}, \textbf{16}(12), 2052--2067.
\newblock 10.1029/2018SW002003.

\bibitem[{{Edwards} et~al.(2022){Edwards}, {Kuridze}, {Williams}, and
  {Morgan}}]{edwards2021}
{Edwards}, L., D.~{Kuridze}, T.~{Williams}, and H.~{Morgan}, 2022.
\newblock A Solar-cycle Study of Coronal Rotation: Large Variations, Rapid
  Changes, and Implications for Solar-wind Models.
\newblock \emph{The Astrophysical Journal}, \textbf{928}(1), 42.
\newblock 10.3847/1538-4357/ac54ba.

\bibitem[{Franses and Wiemann(2020)}]{Franses2020}
Franses, P.~H., and T.~Wiemann, 2020.
\newblock {Intertemporal Similarity of Economic Time Series: An Application of
  Dynamic Time Warping}.
\newblock \emph{Computational Economics}.
\newblock 10.1007/s10614-020-09986-0.

\bibitem[{{Frazin}(2000)}]{frazin2000}
{Frazin}, R.~A., 2000.
\newblock {Tomography of the Solar Corona. I. A Robust, Regularized, Positive
  Estimation Method}.
\newblock \emph{\apj}, \textbf{530}(2), 1026--1035.
\newblock 10.1086/308412.

\bibitem[{{Gonzi} et~al.(2021){Gonzi}, {Weinzierl}, {Bocquet}, {Bisi},
  {Odstrcil}, {Jackson}, {Yeates}, {Jackson}, {Henney}, and {Nick
  Arge}}]{gonzi2021}
{Gonzi}, S., M.~{Weinzierl}, F.~X. {Bocquet}, M.~M. {Bisi}, D.~{Odstrcil},
  B.~V. {Jackson}, A.~R. {Yeates}, D.~R. {Jackson}, C.~J. {Henney}, and
  C.~{Nick Arge}, 2021.
\newblock {Impact of Inner Heliospheric Boundary Conditions on Solar Wind
  Predictions at Earth}.
\newblock \emph{Space Weather}, \textbf{19}(1), e02499.
\newblock 10.1029/2020SW002499.

\bibitem[{{Habbal} et~al.(1997){Habbal}, {Woo}, {Fineschi}, {O'Neal}, {Kohl},
  {Noci}, and {Korendyke}}]{habbal1997}
{Habbal}, S.~R., R.~{Woo}, S.~{Fineschi}, R.~{O'Neal}, J.~{Kohl}, G.~{Noci},
  and C.~{Korendyke}, 1997.
\newblock {Origins of the Slow and the Ubiquitous Fast Solar Wind}.
\newblock \emph{\apjl}, \textbf{489}(1), L103--L106.
\newblock 10.1086/310970, \eprint{astro-ph/9709021}.

\bibitem[{Hinterreiter et~al.(2021)Hinterreiter, Amerstorfer, Temmer, Reiss,
  Weiss, Möstl, Barnard, Pomoell, Bauer, and Amerstorfer}]{hinterreiter2021}
Hinterreiter, J., T.~Amerstorfer, M.~Temmer, M.~A. Reiss, A.~J. Weiss,
  C.~Möstl, L.~A. Barnard, J.~Pomoell, M.~Bauer, and U.~V. Amerstorfer, 2021.
\newblock Drag-based CME modeling with heliospheric images incorporating
  frontal deformation: ELEvoHI 2.0.
\newblock \eprint{2108.08075}.

\bibitem[{{Howard} et~al.(2008){Howard}, {Moses}, {Vourlidas}, {Newmark},
  {Socker} et~al.}]{howard2002}
{Howard}, R.~A., J.~D. {Moses}, A.~{Vourlidas}, J.~S. {Newmark}, D.~G.
  {Socker}, et~al., 2008.
\newblock {Sun Earth Connection Coronal and Heliospheric Investigation
  (SECCHI)}.
\newblock \emph{\ssr}, \textbf{136}(1-4), 67--115.
\newblock 10.1007/s11214-008-9341-4.

\bibitem[{{Imken} et~al.(2018){Imken}, {Randolph}, {DiNicola}, and
  {Nicholas}}]{Imken2018}
{Imken}, T., T.~{Randolph}, M.~{DiNicola}, and A.~{Nicholas}, 2018.
\newblock Modeling spacecraft safe mode events.
\newblock In 2018 IEEE Aerospace Conference, 1--13.
\newblock 10.1109/AERO.2018.8396383.

\bibitem[{Jackson et~al.(2020)Jackson, Buffington, Cota, Odstrcil, Bisi,
  Fallows, and Tokumaru}]{Jackson2020}
Jackson, B.~V., A.~Buffington, L.~Cota, D.~Odstrcil, M.~M. Bisi, R.~Fallows,
  and M.~Tokumaru, 2020.
\newblock Iterative Tomography: A Key to Providing Time-Dependent 3-D
  Reconstructions of the Inner Heliosphere and the Unification of Space Weather
  Forecasting Techniques.
\newblock \emph{Frontiers in Astronomy and Space Sciences}, \textbf{7}.
\newblock 10.3389/fspas.2020.568429,
  \urlprefix\url{https://www.frontiersin.org/article/10.3389/fspas.2020.568429}.

\bibitem[{Jackson et~al.(2010)Jackson, Buffington, Hick, Clover, Bisi, and
  Webb}]{Jackson_2010}
Jackson, B.~V., A.~Buffington, P.~P. Hick, J.~M. Clover, M.~M. Bisi, and D.~F.
  Webb, 2010.
\newblock SMEI 3D reconstruction of a coronal mass ejection interacting with a
  corotating solar wind density enhancement: the 2008 April CME.
\newblock \emph{The Astrophysical Journal}, \textbf{724}(2), 829--834.
\newblock 10.1088/0004-637x/724/2/829.

\bibitem[{{Jackson} et~al.(2013){Jackson}, {Clover}, {Hick}, {Buffington},
  {Bisi}, and {Tokumaru}}]{Jackson2013}
{Jackson}, B.~V., J.~M. {Clover}, P.~P. {Hick}, A.~{Buffington}, M.~M. {Bisi},
  and M.~{Tokumaru}, 2013.
\newblock {Inclusion of Real-Time In-Situ Measurements into the UCSD
  Time-Dependent Tomography and Its Use as a Forecast Algorithm}.
\newblock \emph{\solphys}, \textbf{285}(1-2), 151--165.
\newblock 10.1007/s11207-012-0102-x.

\bibitem[{{Jang} et~al.(2021){Jang}, {Kwon}, {Linker}, {Riley}, {Shin},
  {Downs}, and {Kim}}]{jang2021}
{Jang}, S., R.-Y. {Kwon}, J.~A. {Linker}, P.~{Riley}, G.~{Shin}, C.~{Downs},
  and Y.-H. {Kim}, 2021.
\newblock {Development of a Deep Learning Model for Inversion of Rotational
  Coronagraphic Images Into 3D Electron Density}.
\newblock \emph{\apjl}, \textbf{920}(2), L30.
\newblock 10.3847/2041-8213/ac2a46.

\bibitem[{{Kaiser}(2005)}]{kaiser2005}
{Kaiser}, M.~L., 2005.
\newblock {The STEREO mission: an overview}.
\newblock \emph{Advances in Space Research}, \textbf{36}, 1483--1488.
\newblock 10.1016/j.asr.2004.12.066.

\bibitem[{Kasper et~al.(2019)Kasper, Bale, Belcher, Berthomier, Case
  et~al.}]{Kasper2019}
Kasper, J.~C., S.~D. Bale, J.~W. Belcher, M.~Berthomier, A.~W. Case, et~al.,
  2019.
\newblock {Alfv{\'{e}}nic velocity spikes and rotational flows in the near-Sun
  solar wind}.
\newblock \emph{Nature}.
\newblock 10.1038/s41586-019-1813-z.

\bibitem[{{Kramar} et~al.(2014){Kramar}, {Airapetian}, {Miki{\'c}}, and
  {Davila}}]{kramar2014}
{Kramar}, M., V.~{Airapetian}, Z.~{Miki{\'c}}, and J.~{Davila}, 2014.
\newblock {3D Coronal Density Reconstruction and Retrieving the Magnetic Field
  Structure during Solar Minimum}.
\newblock \emph{\solphys}, \textbf{289}(8), 2927--2944.
\newblock 10.1007/s11207-014-0525-7, \eprint{1405.0951}.

\bibitem[{Linker et~al.(1999)Linker, Mikić, Biesecker, Forsyth, Gibson,
  Lazarus, Lecinski, Riley, Szabo, and Thompson}]{Linker1999}
Linker, J.~A., Z.~Mikić, D.~A. Biesecker, R.~J. Forsyth, S.~E. Gibson, A.~J.
  Lazarus, A.~Lecinski, P.~Riley, A.~Szabo, and B.~J. Thompson, 1999.
\newblock Magnetohydrodynamic modeling of the solar corona during Whole Sun
  Month.
\newblock \emph{Journal of Geophysical Research: Space Physics}, \textbf{104}.
\newblock 10.1029/1998ja900159.

\bibitem[{MacNeice et~al.(2018)MacNeice, Jian, Antiochos, Arge, Bussy-Virat
  et~al.}]{macneice2018}
MacNeice, P., L.~K. Jian, S.~K. Antiochos, C.~N. Arge, C.~D. Bussy-Virat,
  et~al., 2018.
\newblock Assessing the Quality of Models of the Ambient Solar Wind.
\newblock \emph{Space Weather}, \textbf{16}(11), 1644--1667.
\newblock Https://doi.org/10.1029/2018SW002040.

\bibitem[{Meziane et~al.(2014)Meziane, Alrefay, and Hamza}]{MEZIANE20141}
Meziane, K., T.~Alrefay, and A.~Hamza, 2014.
\newblock On the shape and motion of the Earth's bow shock.
\newblock \emph{Planetary and Space Science}, \textbf{93-94}, 1--9.
\newblock Https://doi.org/10.1016/j.pss.2014.01.006.

\bibitem[{Milan et~al.(2007)Milan, Provan, and Hubert}]{Milan2007}
Milan, S.~E., G.~Provan, and B.~Hubert, 2007.
\newblock {Magnetic flux transport in the Dungey cycle: A survey of dayside and
  nightside reconnection rates}.
\newblock \emph{Journal of Geophysical Research: Space Physics}.
\newblock 10.1029/2006JA011642.

\bibitem[{{Morgan}(2011)}]{morgan2011rotation}
{Morgan}, H., 2011.
\newblock {The Rotation of the White Light Solar Corona at Height 4 R $_{sun}$
  from 1996 to 2010: A Tomographical Study of Large Angle and Spectrometric
  Coronagraph C2 Observations}.
\newblock \emph{\apj}, \textbf{738}(2), 189.
\newblock 10.1088/0004-637X/738/2/189.

\bibitem[{Morgan(2011a)}]{Morgan2011}
Morgan, H., 2011a.
\newblock {The rotation of the white light solar corona at height 4
  R$_{{\ensuremath{\odot}}}$ from 1996 to 2010: A tomographical study of large
  angle and spectrometric coronagraph C2 observations}.
\newblock \emph{Astrophysical Journal}.
\newblock 10.1088/0004-637X/738/2/189.

\bibitem[{{Morgan}(2011b)}]{morgan2011b}
{Morgan}, H., 2011b.
\newblock Longitudinal Drifts of Streamers across the Heliospheric Current
  Sheet.
\newblock \emph{The Astrophysical Journal}, \textbf{738}(2), 190.
\newblock 10.1088/0004-637x/738/2/190.

\bibitem[{{Morgan}(2015)}]{morgan2015}
{Morgan}, H., 2015.
\newblock {An Atlas of Coronal Electron Density at 5R$_{{\ensuremath{\odot}}}$.
  I. Data Processing and Calibration}.
\newblock \emph{\apjs}, \textbf{219}(2), 23.
\newblock 10.1088/0067-0049/219/2/23, \eprint{1509.03113}.

\bibitem[{Morgan(2019)}]{Morgan2019}
Morgan, H., 2019.
\newblock {An atlas of coronal electron density at 5R$_{{\ensuremath{\odot}}}$.
  II: A spherical harmonic method for density reconstruction}.
\newblock 10.3847/1538-4365/ab125d.

\bibitem[{{Morgan}(2021)}]{morgan2021}
{Morgan}, H., 2021.
\newblock {Daily Variations of Plasma Density in the Solar Streamer Belt}.
\newblock \emph{\apj}, \textbf{922}(2), 165.
\newblock 10.3847/1538-4357/ac1799.

\bibitem[{{Morgan} et~al.(2012){Morgan}, {Byrne}, and {Habbal}}]{morgan2012cme}
{Morgan}, H., J.~P. {Byrne}, and S.~R. {Habbal}, 2012.
\newblock {Automatically Detecting and Tracking Coronal Mass Ejections. I.
  Separation of Dynamic and Quiescent Components in Coronagraph Images}.
\newblock \emph{\apj}, \textbf{752}, 144.
\newblock 10.1088/0004-637X/752/2/144.

\bibitem[{{Morgan} and {Cook}(2020)}]{morgan2020}
{Morgan}, H., and A.~C. {Cook}, 2020.
\newblock {The Width, Density, and Outflow of Solar Coronal Streamers}.
\newblock \emph{\apj}, \textbf{893}(1), 57.
\newblock 10.3847/1538-4357/ab7e32, \eprint{2003.04809}.

\bibitem[{{Morgan} and {Habbal}(2007{\natexlab{a}})}]{morgan2007empiricalmodel}
{Morgan}, H., and S.~R. {Habbal}, 2007{\natexlab{a}}.
\newblock {An empirical 3D model of the large-scale coronal structure based on
  the distribution of H{\ensuremath{\alpha}} filaments on the solar disk}.
\newblock \emph{\aap}, \textbf{464}(1), 357--365.
\newblock 10.1051/0004-6361:20066482, \eprint{astro-ph/0610219}.

\bibitem[{{Morgan} and {Habbal}(2007{\natexlab{b}})}]{morgan2007twistingsheets}
{Morgan}, H., and S.~R. {Habbal}, 2007{\natexlab{b}}.
\newblock {Are solar maximum fan streamers a consequence of twisting sheet
  structures?}
\newblock \emph{\aap}, \textbf{465}(3), L47--L50.
\newblock 10.1051/0004-6361:20077126.

\bibitem[{{Morgan} and {Habbal}(2010{\natexlab{a}})}]{morgan2010structure}
{Morgan}, H., and S.~R. {Habbal}, 2010{\natexlab{a}}.
\newblock {Observational Aspects of the Three-dimensional Coronal Structure
  Over a Solar Activity Cycle}.
\newblock \emph{\apj}, \textbf{710}(1), 1--15.
\newblock 10.1088/0004-637X/710/1/1.

\bibitem[{{Morgan} and {Habbal}(2010{\natexlab{b}})}]{morgan2010}
{Morgan}, H., and S.~R. {Habbal}, 2010{\natexlab{b}}.
\newblock {Observational Aspects of the Three-dimensional Coronal Structure
  Over a Solar Activity Cycle}.
\newblock \emph{\apj}, \textbf{710}(1), 1--15.
\newblock 10.1088/0004-637X/710/1/1.

\bibitem[{{Morgan} et~al.(2009){Morgan}, {Habbal}, and {Lugaz}}]{morgan2009}
{Morgan}, H., S.~R. {Habbal}, and N.~{Lugaz}, 2009.
\newblock {Mapping the Structure of the Corona Using Fourier Backprojection
  Tomography}.
\newblock \emph{\apj}, \textbf{690}(2), 1119--1129.
\newblock 10.1088/0004-637X/690/2/1119.

\bibitem[{{Morgan, H.} and {Habbal, S. R.}(2007)}]{morgan2007fcorona}
{Morgan, H.}, and {Habbal, S. R.}, 2007.
\newblock The long-term stability of the visible F corona at heights of 3-6
  R$_{{\ensuremath{\odot}}}$.
\newblock \emph{A\&A}, \textbf{471}(2), L47--L50.
\newblock 10.1051/0004-6361:20078071.

\bibitem[{Odstrcil(2003)}]{ODSTRCIL2003497}
Odstrcil, D., 2003.
\newblock Modeling 3-D solar wind structure.
\newblock \emph{Advances in Space Research}, \textbf{32}(4), 497--506.
\newblock Heliosphere at Solar Maximum,
  https://doi.org/10.1016/S0273-1177(03)00332-6.

\bibitem[{Odstrcil et~al.(2004)Odstrcil, Riley, and Zhao}]{Odstrcil2004}
Odstrcil, D., P.~Riley, and X.~P. Zhao, 2004.
\newblock Numerical simulation of the 12 May 1997 interplanetary CME event.
\newblock \emph{Journal of Geophysical Research: Space Physics},
  \textbf{109}(A2).
\newblock Https://doi.org/10.1029/2003JA010135.

\bibitem[{{Owens}(2018)}]{owens2018}
{Owens}, M.~J., 2018.
\newblock {Time-Window Approaches to Space-Weather Forecast Metrics: A Solar
  Wind Case Study}.
\newblock \emph{Space Weather}, \textbf{16}(11), 1847--1861.
\newblock 10.1029/2018SW002059.

\bibitem[{Owens et~al.(2013)Owens, Challen, Methven, Henley, and
  Jackson}]{owens2013}
Owens, M.~J., R.~Challen, J.~Methven, E.~Henley, and D.~R. Jackson, 2013.
\newblock A 27 day persistence model of near-Earth solar wind conditions: A
  long lead-time forecast and a benchmark for dynamical models.
\newblock \emph{Space Weather}, \textbf{11}(5), 225--236.
\newblock 10.1002/swe.20040.

\bibitem[{Owens et~al.(2020)Owens, Lang, Barnard, Riley, Ben-Nun, Scott,
  Lockwood, Reiss, Arge, and Gonzi}]{Owens2020}
Owens, M.~J., M.~Lang, L.~Barnard, P.~Riley, M.~Ben-Nun, C.~J. Scott,
  M.~Lockwood, M.~A. Reiss, C.~N. Arge, and S.~Gonzi, 2020.
\newblock {A Computationally Efficient, Time-Dependent Model of the Solar Wind
  for Use as a Surrogate to Three-Dimensional Numerical Magnetohydrodynamic
  Simulations}.
\newblock \emph{Solar Physics}.
\newblock 10.1007/s11207-020-01605-3.

\bibitem[{Owens and Nichols(2021)}]{Owens_2021}
Owens, M.~J., and J.~D. Nichols, 2021.
\newblock Using in-situ solar-wind observations to generate inner-boundary
  conditions to outer-heliosphere simulations, 1: Dynamic time warping applied
  to synthetic observations.
\newblock \emph{Monthly Notices of the Royal Astronomical Society}.
\newblock 10.1093/mnras/stab2512.

\bibitem[{Owens and Riley(2017)}]{Owens_Riley_2017}
Owens, M.~J., and P.~Riley, 2017.
\newblock Probabilistic Solar Wind Forecasting Using Large Ensembles of
  Near-Sun Conditions With a Simple One-Dimensional “Upwind” Scheme.
\newblock \emph{Space Weather}, \textbf{15}(11), 1461--1474.
\newblock 10.1002/2017SW001679.

\bibitem[{Parker et~al.(1964)Parker, Marshak, and Johnson}]{Parker1964}
Parker, E.~N., R.~E. Marshak, and G.~Johnson, 1964.
\newblock Interplanetary Dynamical Processes.
\newblock \emph{Physics Today}, \textbf{17}.
\newblock 10.1063/1.3051487.

\bibitem[{{Poirier} et~al.(2020){Poirier}, {Kouloumvakos}, {Rouillard},
  {Pinto}, {Vourlidas} et~al.}]{poirier2020}
{Poirier}, N., A.~{Kouloumvakos}, A.~P. {Rouillard}, R.~F. {Pinto},
  A.~{Vourlidas}, et~al., 2020.
\newblock {Detailed Imaging of Coronal Rays with the Parker Solar Probe}.
\newblock \emph{\apjs}, \textbf{246}(2), 60.
\newblock 10.3847/1538-4365/ab6324, \eprint{1912.09345}.

\bibitem[{Poirier et~al.(2021)Poirier, Rouillard, Kouloumvakos, Przybylak,
  Fargette, Pobeda, Réville, Pinto, Indurain, and Alexandre}]{poirier2021}
Poirier, N., A.~P. Rouillard, A.~Kouloumvakos, A.~Przybylak, N.~Fargette,
  R.~Pobeda, V.~Réville, R.~F. Pinto, M.~Indurain, and M.~Alexandre, 2021.
\newblock Exploiting White-Light Observations to Improve Estimates of Magnetic
  Connectivity.
\newblock \emph{Frontiers in Astronomy and Space Sciences}, \textbf{8}, 84.
\newblock 10.3389/fspas.2021.684734.

\bibitem[{{Pomoell, J.} and {Poedts, S.}(2018)}]{Pomell2018}
{Pomoell, J.}, and {Poedts, S.}, 2018.
\newblock EUHFORIA: European heliospheric forecasting information asset.
\newblock \emph{J. Space Weather Space Clim.}, \textbf{8}, A35.
\newblock 10.1051/swsc/2018020.

\bibitem[{Riley et~al.(2012)Riley, Linker, Lionello, and Mikic}]{Riley2012}
Riley, P., J.~A. Linker, R.~Lionello, and Z.~Mikic, 2012.
\newblock Corotating interaction regions during the recent solar minimum: The
  power and limitations of global MHD modeling.
\newblock \emph{Journal of Atmospheric and Solar-Terrestrial Physics},
  \textbf{83}.
\newblock 10.1016/j.jastp.2011.12.013.

\bibitem[{Riley and Lionello(2011)}]{Riley2011}
Riley, P., and R.~Lionello, 2011.
\newblock {Mapping Solar Wind Streams from the Sun to 1 AU: A Comparison of
  Techniques}.
\newblock \emph{Solar Physics}.
\newblock 10.1007/s11207-011-9766-x.

\bibitem[{Salvador and Chan(2004)}]{Salvador2004}
Salvador, S., and P.~Chan, 2004.
\newblock {FastDTW: Toward Accurate Dynamic Time Warping in Linear Time and
  Space}.
\newblock In KDD workshop on mining temporal and sequential data.

\bibitem[{Samara et~al.(2022)Samara, Laperre, Kieokaew, Temmer, Verbeke,
  Rodriguez, Magdaleni{\'{c}}, and Poedts}]{Samara2021}
Samara, E., B.~Laperre, R.~Kieokaew, M.~Temmer, C.~Verbeke, L.~Rodriguez,
  J.~Magdaleni{\'{c}}, and S.~Poedts, 2022.
\newblock Dynamic Time Warping as a Means of Assessing Solar Wind Time Series.
\newblock \emph{The Astrophysical Journal}, \textbf{927}(2), 187.
\newblock 10.3847/1538-4357/ac4af6,
  \urlprefix\url{https://doi.org/10.3847/1538-4357/ac4af6}.

\bibitem[{Schwenn(1990)}]{Schwenn1990}
Schwenn, R., 1990.
\newblock {Large-Scale Structure of the Interplanetary Medium}.
\newblock In Physics of the Inner Heliosphere I. Springer, Berlin, Heidelberg.
\newblock Https://doi.org/10.1007/978-3-642-75361-9.

\bibitem[{Schwenn(2006)}]{Schwenn2006}
Schwenn, R., 2006.
\newblock Space Weather: The Solar Perspective.
\newblock \emph{Living Reviews in Solar Physics}, \textbf{3}.
\newblock 10.12942/lrsp-2006-2.

\bibitem[{Skutkova et~al.(2013)Skutkova, Vitek, Babula, Kizek, and
  Provaznik}]{Skutkova2013}
Skutkova, H., M.~Vitek, P.~Babula, R.~Kizek, and I.~Provaznik, 2013.
\newblock {Classification of genomic signals using dynamic time warping}.
\newblock \emph{BMC Bioinformatics}.
\newblock 10.1186/1471-2105-14-S10-S1.

\bibitem[{{Thernisien} and {Howard}(2006)}]{thernisien2006}
{Thernisien}, A.~F., and R.~A. {Howard}, 2006.
\newblock {Electron Density Modeling of a Streamer Using LASCO Data of 2004
  January and February}.
\newblock \emph{\apj}, \textbf{642}, 523--532.
\newblock 10.1086/500818.

\bibitem[{{Vibert} et~al.(2016){Vibert}, {Peillon}, {Lamy}, {Frazin}, and
  {Wojak}}]{vibert2016}
{Vibert}, D., C.~{Peillon}, P.~{Lamy}, R.~A. {Frazin}, and J.~{Wojak}, 2016.
\newblock {Time-dependent tomographic reconstruction of the solar corona}.
\newblock \emph{Astronomy and Computing}, \textbf{17}, 144--162.
\newblock 10.1016/j.ascom.2016.09.001, \eprint{1607.06308}.

\bibitem[{{Wang} and {Sheeley}(1990)}]{WangSheeley1990}
{Wang}, Y.~M., and J.~{Sheeley}, N.~R., 1990.
\newblock {Solar Wind Speed and Coronal Flux-Tube Expansion}.
\newblock \emph{\apj}, \textbf{355}, 726.
\newblock 10.1086/168805.

\bibitem[{{Weinzierl} et~al.(2016){Weinzierl}, {Yeates}, {Mackay}, {Henney},
  and {Arge}}]{weinzierl2016}
{Weinzierl}, M., A.~R. {Yeates}, D.~H. {Mackay}, C.~J. {Henney}, and C.~N.
  {Arge}, 2016.
\newblock {A New Technique for the Photospheric Driving of Non-potential Solar
  Coronal Magnetic Field Simulations}.
\newblock \emph{\apj}, \textbf{823}(1), 55.
\newblock 10.3847/0004-637X/823/1/55.

\bibitem[{{Yeates} et~al.(2018){Yeates}, {Amari}, {Contopoulos}, {Feng},
  {Mackay} et~al.}]{yeates2018}
{Yeates}, A.~R., T.~{Amari}, I.~{Contopoulos}, X.~{Feng}, D.~H. {Mackay},
  et~al., 2018.
\newblock {Global Non-Potential Magnetic Models of the Solar Corona During the
  March 2015 Eclipse}.
\newblock \emph{\ssr}, \textbf{214}(5), 99.
\newblock 10.1007/s11214-018-0534-1, \eprint{1808.00785}.

\bibitem[{{Yeates} et~al.(2008){Yeates}, {Mackay}, and {van
  Ballegooijen}}]{yeates2008}
{Yeates}, A.~R., D.~H. {Mackay}, and A.~A. {van Ballegooijen}, 2008.
\newblock {Modelling the Global Solar Corona II: Coronal Evolution and Filament
  Chirality Comparison}.
\newblock \emph{\solphys}, \textbf{247}(1), 103--121.
\newblock 10.1007/s11207-007-9097-0, \eprint{0711.2887}.

\end{thebibliography}

\end{document}